\documentclass[a4paper,12pt]{article}
\pdfoutput=1
\usepackage{rotating}

\usepackage{ucs}
\usepackage{amsmath}
\usepackage{amsfonts}
\usepackage{amssymb}
\usepackage{eucal}
\usepackage{mathrsfs}          
\usepackage[sans]{dsfont}     
\usepackage{slashed}
\usepackage[Symbol]{upgreek}  
\usepackage{array}            
\usepackage{cite}               
\usepackage[normalem]{ulem}             
\usepackage{booktabs}      
\usepackage{colortbl}         
\usepackage{fancyhdr}
\usepackage{multirow}
\usepackage{feynmf}           
\usepackage
{graphicx}

\ifx\pdfoutput\undefined
\usepackage[dvips,bookmarks]{hyperref}	
\else
\usepackage{hyperref}	
\fi
\hypersetup{colorlinks,bookmarksopen,bookmarksnumbered,citecolor=verdes,
linkcolor=blus,pdfstartview=FitH,urlcolor=rossos}

\def\hhref#1{\href{http://arxiv.org/abs/#1}{#1}} 

\usepackage{multicol}
\usepackage{color}
\definecolor{rosso}{cmyk}{0,1,1,0.4}
\definecolor{rossos}{cmyk}{0,1,1,0.55}
\definecolor{rossoc}{cmyk}{0,1,1,0.2}
\definecolor{blu}{cmyk}{1,1,0,0.3}
\definecolor{blus}{cmyk}{1,1,0,0.6}
\definecolor{bluc}{cmyk}{1,1,0,0.1}
\definecolor{verde}{cmyk}{0.92,0,0.59,0.25}
\definecolor{verdec}{cmyk}{0.92,0,0.59,0.15}
\definecolor{verdes}{cmyk}{0.92,0,0.59,0.4}

\font\tenrsfs=rsfs10 at 12pt
\font\sevenrsfs=rsfs7
\font\fiversfs=rsfs5
\newfam\rsfsfam
\textfont\rsfsfam=\tenrsfs
\scriptfont\rsfsfam=\sevenrsfs
\scriptscriptfont\rsfsfam=\fiversfs
\def\mathscr#1{{\fam\rsfsfam\relax#1}}
\def\Lag{\mathscr{L}}

\def\circa#1{\,\raise.3ex\hbox{$#1$\kern-.75em\lower1ex\hbox{$\sim$}}\,}

\newcommand{\DAMA}{{\sf DAMA}}
\newcommand{\DAMANaI}{{\sf DAMA/NaI}}
\newcommand{\DAMALibra}{{\sf DAMA/Libra}}

\newcommand{\CoGeNT}{{\sf CoGeNT}}

\newcommand{\CRESST}{{\sf CRESST-II}}

\newcommand{\CDMS}{{\sf CDMS}}
\newcommand{\CDMSSi}{{\sf CDMS-Si}}
\newcommand{\CDMSGe}{{\sf CDMS-Ge}}

\newcommand{\XENON}{{\sf XENON}}
\newcommand{\XENONten}{{\sf XENON10}}
\newcommand{\XENONhundred}{{\sf XENON100}}

\newcommand{\COUPP}{{\sf COUPP}}

\newcommand{\PICASSO}{{\sf PICASSO}}

\newcommand{\LUX}{{\sf LUX}}

\newcommand{\LHC}{{\sf LHC}}

\newcommand{\Op}{{\cal O}}		
\newcommand{\Mel}{\CMcal{M}}	
\font\bb=bbmss10 scaled 1200
\newcommand{\unop}{\mbox{\bb 1}}
\newcommand{\ud}{\text{d}}

\newcommand{\ER}{E_\text{R}}
\newcommand{\Ed}{E'}
\newcommand{\vesc}{v_\text{esc}}
\newcommand{\vmin}{v_\text{min}}

\newcommand{\NR}{\text{NR}}

\newcommand{\beq}{\begin{equation}}
\newcommand{\eeq}{\end{equation}}

\def\circa#1{\,\raise.3ex\hbox{$#1$\kern-.75em\lower1ex\hbox{$\sim$}}\,}
\makeatletter

\def\art{\@ifnextchar[{\eart}{\oart}}
\def\eart[#1]#2#3#4#5#6{{\rm #2}, {#3 #4} {\rm (#6) #5} [{\hhref{#1}}]}
\def\hepart[#1]#2{{\rm #2, \hhref{#1}}}
\newcommand{\oart}[5]{{\rm #1}, {#2 #3} {\rm (#5) #4}}

\newcounter{alphaequation}[equation]
\def\thealphaequation{\theequation\hbox to
0.6em{\hfil\alph{alphaequation}\hfil}}
\def\eqnsystem#1{
\def\@eqnnum{{\rm (\thealphaequation)}}
\def\@@eqncr{\let\@tempa\relax \ifcase\@eqcnt \def\@tempa{& & &} \or
  \def\@tempa{& &}\or \def\@tempa{&}\fi\@tempa
  \if@eqnsw\@eqnnum\refstepcounter{alphaequation}\fi
\global\@eqnswtrue\global\@eqcnt=0\cr}
\refstepcounter{equation} \let\@currentlabel\theequation \def\@tempb{#1}
\ifx\@tempb\empty\else\label{#1}\fi
\refstepcounter{alphaequation}
\let\@currentlabel\thealphaequation
\global\@eqnswtrue\global\@eqcnt=0 \tabskip\@centering\let\\=\@eqncr
$$\halign to \displaywidth\bgroup \@eqnsel\hskip\@centering
$\displaystyle\tabskip\z@{##}$&\global\@eqcnt\@ne
\hskip2\arraycolsep\hfil${##}$\hfil& \global\@eqcnt\tw@\hskip2\arraycolsep
$\displaystyle\tabskip\z@{##}$\hfil
\tabskip\@centering&\llap{##}\tabskip\z@\cr}
\def\endeqnsystem{\@@eqncr\egroup$$\global\@ignoretrue} \makeatother

\begin{document}
\begin{flushright}
\footnotesize
\end{flushright}
\color{black}

\begin{center}
{\Huge\bf New Directions in Direct \\ \vspace{.25cm}  Dark Matter Searches}

\medskip
\bigskip\color{black}\vspace{0.6cm}

{
{\large\bf Paolo Panci}\ $^a$,
}
\\[7mm]
{\it $^a$ \href{http://iap.fr}{Institut d'Astrophysique}, CNRS, UMR 7095 \& Universit\'e Pierre et Marie Curie,\\ 
	98bis Boulevard Arago, F-75014 Paris, France}\\[3mm]
\end{center}

\bigskip

\centerline{\large\bf Abstract}
\begin{quote}
\color{black}\large
I  present the status of direct dark matter detection with  specific attention to the experimental results and their phenomenological interpretation in terms of dark matter interactions. In particular I review a new and more general approach to study signals in this field based on non-relativistic operators which parametrize more efficiently the dark matter-nucleus interactions in terms of a very limited number of relevant degrees of freedom. Then I list the major experimental results, pointing out the main uncertainties that affect the theoretical interpretation of the data.  Finally, since the underlying theory that describes both the dark matter and the standard model fields is unknown, I address the uncertainties coming from the nature of the interaction. In particular, the phenomenology of a class of models in which the interaction between dark matter particles and target nuclei is of a long-range type is discussed. 
\end{quote}

\newpage

\tableofcontents

\section{Introduction}

One of the most exciting open questions at the interface between particle physics and cosmology is the nature of Dark Matter (DM). The first person who provided evidence and inferred the presence of DM was a Swiss-American astrophysicist, Fritz Zwicky. He applied the virial theorem to the Coma cluster of galaxies and obtained evidence of unseen mass. Roughly 40 years following the discoveries of Zwicky and others, Vera Rubin and collaborators conducted an extensive study of the rotation curves of isolated spiral galaxies. They announced the discovery that the rotational curves of stars in spiral galaxies exhibit a characteristic flat behaviour at large distance in contrast with Kepler's law. Many other evidence of unseen mass on distance scales of the size of galaxies and clusters of galaxies appeared throughout the years, but the most precise measurement of the total amount of DM comes from cosmological scales. In particular, the measurements of modern precision cosmology (the Cosmic Microwave Background (CMB) and the surveys of the Large Scale Structure (LSS) of the Universe), provide the current most relevant evidence. Apart from the qualitative agreement, it is the quantitative fitting of the wealth of available data that allows the amount of DM to be one of the cosmological parameters now most precisely measured ($\Omega_\chi h^2=0.1199\pm0.0027$, see Tab.~2 of~\cite{Ade:2013zuv}).  Therefore we have compelling evidence of unseen mass, but the microscopic features of this new kind of matter remain unknown yet. Direct and Indirect searches may shed light on the nature of DM, and therefore a careful study of their phenomenology is fundamental. For a pedagogical review on this subject, see e.g.~\cite{Bertone:2004pz}. 

\medskip
Direct searches for DM aim at detecting the nuclear recoils arising from scattering between DM particles and target nuclei in underground detectors. DM direct detection experiments are providing exciting results in terms of measured features which have the right properties to be potentially ascribed to a DM signal.  For example in addition to the long-standing \DAMA\ results, nowadays there are other experiments, such as \CoGeNT, \CRESST\ and \CDMSSi\, that start to see some anomalies in their counting rates. On the other hand ,the situation in this field is extremely unclear and confusing, because on top on these positive result experiments, the constraints coming from null results, like \XENONhundred, \COUPP, \PICASSO\ and very recently \LUX, are very stringent and put  the interpretation of the anomalies in terms of a DM interaction in serious trouble. Nevertheless, there are at least two main caveats when the results from the experiments commented upon above are interpreted. The first is that one has to treat with great care the fine experimental details associated with the results quoted by each experiment. The second caveat is instead associated with the interpretation of the data within a very simple-minded DM model. For instance, the DM-nucleus spin independent contact interaction is just a benchmark example. Upon relaxing some of these assumptions, the current complicated experimental puzzle can probably be solved. 

\medskip
The {\bf scope of this work} is to present the {\em status} of direct DM detection with specific attention to the experimental results and their phenomenological interpretation in terms of DM interaction. In particular in Sec.~\ref{Basics} I review a new and more general approach to study signals in this field based on non-relativistic operators which parameterize most efficiently the DM-nucleus interactions in terms of a very limited number of relevant degrees of freedom. In Sec.~\ref{DDStatus} I review the experimental results and their interpretation in terms of the ``standard'' spin independent (SI) interaction. I list then the main uncertainties that affect the theoretical interpretation of the data: this is a very promising area of research since only major advancements here can probably reconcile the complicated puzzle showed  by the experiments up. Finally in Sec.~\ref{LRInteraction}, I pose my attention on the uncertainties coming from the nature of the interaction. In particular the phenomenology  of a class of models in which the interaction between DM particles and target nuclei is of a long range type is discussed.

\section{Basics of Direct Detection computations}\label{Basics}

\subsection{Kinematics}
As already stated, when a DM particle scatters off a nucleus, depending on the DM properties, one can envision at least two distinct kinematics, the elastic and the inelastic one. The elastic scattering is represented by
\beq
\chi+ {\cal N}(A_{\mathcal N},Z_{\mathcal N})_{\rm at \,rest}\rightarrow \chi+ {\cal N}(A_{\mathcal N},Z_{\mathcal N})_{\rm recoil} \ ,
\label{eq:elastic}
\eeq
while the inelastic is
\beq
\chi+ {\cal N}(A_{\mathcal N},Z_{\mathcal N})_{\rm at \,rest}\rightarrow \chi'+ {\cal N}(A_{\mathcal N},Z_{\mathcal N})_{\rm recoil} \ .
\label{eq:inelastic}
\eeq
In (\ref{eq:elastic}) and (\ref{eq:inelastic}), $\chi$ and $\chi'$ are two DM particle states, and $A_{\mathcal N}$, $Z_{\mathcal N}$ are respectively the mass and atomic numbers of the nucleus 
$\cal N$. 

\medskip
We know, thanks to rotational curves data, that the velocity of $\chi$ in the vicinity of the Earth is of the order of 10$^{-3}c$. Hence the scattering between a DM particle with velocity $v$ and mass $m_\chi$ and a nucleus at rest with mass $m_{\cal N}$ occurs in deeply non relativistic regime. This is analogous with the collision between two billiard balls and therefore the recoil energy can be simply obtained by considering energy and momenta conservation separately.  In the detector rest frame, such scattering would end up into a nuclear recoil of energy
\beq\label{RecoilEnergy}
\ER=\frac{2\,\mu_{\chi \cal N}^2}{ m_{\cal N}}v^2\left(\frac{1-\frac{v_{\rm t}^2}{2 v^2}-\sqrt{1-\frac{v_{\rm t}^2}{v^2}}\,\cos\theta}2\right), \qquad 
\left\{\begin{array}{l}
v_{\rm t}=0, \qquad \mbox{Elastic Scattering} \\
v_{\rm t}\neq0, \qquad \mbox{Inelastic Scattering}
\end{array}\right.,
\eeq
where $\mu_{\chi \cal N}=m_\chi m_{\cal N}/(m_\chi+m_{\cal N})$ is the DM-nucleus reduced mass, $\theta$ as usual is the scattering angle in the center-of-mass frame and  $v_{\rm t}=\sqrt{2\delta / \mu_{\chi \cal N} }$ is a threshold velocity. Here $\delta=m_\chi'-m_\chi$ is the mass splitting between $\chi$ and $\chi'$, and the equation above holds for $\delta \ll m_\chi', m_\chi$. Elastic scattering occurs for $\delta = 0$, while $\delta \neq 0$ implies inelastic scattering. The minimal DM velocity providing the $\ER$ recoil energy can be then obtained by putting $\theta=-\pi$ in Eq.~\eqref{RecoilEnergy}. One gets:
\beq
\label{minvelocity}
\vmin(\ER) = \sqrt{ \frac{m_{\cal N}\,\ER}{2\mu_{\chi \cal N}^2} }\left(1+\frac{\mu_{\chi \cal N}\,\delta}{m_{\cal N}\,\ER}\right) \ .
\eeq

\subsection{Formalism of non-relativistic Operators}\label{NRformalism}
Having a disposal the basic quantity $\vmin$ that fully describes the kinematics of the DM-nucleus scatterings in Eqs.~(\ref{eq:elastic},\ref{eq:inelastic}), the differential rate of nuclear recoil expected in a given detector can be achieved by weighting the differential cross section $\ud\sigma_{\cal N}/\ud \ER$ with the DM velocity distribution $f_{\rm E}(\vec v)$ in the velocity range allowed by the kinematics. It reads: 
\beq\label{RN}
\frac{\ud R_{\cal N}}{\ud \ER} = \frac{\xi_{\cal N}}{m_{\cal N}} \frac{\rho_\odot}{m_\chi} \int_{\vmin(\ER)}^{\vesc} \hspace{-.5cm} \ud^3 v \, v \, f_\text{E}(\vec v) \frac{\ud \sigma_{\cal N}}{\ud \ER} (v, \ER) \ ,
\eeq
where $\xi_{\cal N}$ are the mass fractions of different nuclides\footnote{$\xi_{\cal N} = 10^3 N_\text{A} m_{\cal N} \zeta_{\cal N} / {\rm kg} \, \bar{A}$, where $N_\text{A} = 6.022 \times 10^{23}$ is Avogadro's number, $\zeta_{\cal N}$ are the numeric abundances and $\bar{A} \equiv \sum_{\cal N} \zeta_{\cal N} A_{\cal N}$.}, and $\rho_\odot \simeq 0.3$ GeV/cm$^3$ is the DM energy density at the Earth's location. This is the canonical value routinely adopted in the literature (see for instance Ref.~\cite{Bertone:2004pz, Jungman:1995df, Beringer:1900zz}).  Recent computations have found a higher central value of it, still subject for some debate \cite{Catena:2009mf,Weber:2009pt, Salucci:2010qr}. The integral in the right-hand side of the equation above is a key ingredient because it contains all the information related to the geometry of the DM halo, the nature of the DM-nucleus interaction and in turn the nuclear response of the target.   In the following we present the formalism of non-relativistic (NR) operators and then we describe how to write down the main observables in terms of it.

\begin{itemize}
\item As already stated, the weight of the velocity integral is the DM velocity distribution $f_{\rm E}(\vec v)$. In the rest frame of our Galaxy it can be roughly approximated with a Maxwell-Boltzmann distribution due to the fact, as pointed out in Ref.~\cite{LyndenBell}, that the violent relaxation of the gravitational potential at the formation of the Milky Way lead to fast mixing of the DM phase space elements. Therefore DM particles were basically frozen in high-entropy configuration which are indeed Maxwell-Boltzmann like. This has been roughly confirmed by some recent numeral simulation, but as one can see in Fig.~2 of Ref.~\cite{Vogelsberger:2008qb}, there are some deviations due to the DM assembly history of the Milky Way. Indeed one can see different features both at low and high velocities that of course are not present in the case of a Maxwell-Boltzmann distribution. Furthermore the geometry of the halo predicted by this kind of numerical simulation is not exactly spherical but tends to a triaxial configuration (see e.g. Ref.~\cite{Jing:2002np}). 

\medskip
Keeping in mind that this is not the truth because we have not directly measured the properties of the DM halo  yet, what we know for sure is that, since we are setting in the Earth laboratory frame, one has to boost the DM-nucleus relative velocity with the drift velocity of the Sun around the galactic center and the Earth's orbital velocity projected in the galactic plane. Therefore the  DM velocity distribution $f_\text{E}(\vec v)$ in the Earth frame is related to the DM 
velocity distribution in the galactic one $f_\text{G}(\vec w)$ by the galilean velocity transformation $f_\text{E}(\vec v) = f_\text{G}(\vec v + \vec v_\text{E}(t))$, where
\beq\label{vE}
\vec v_\text{E}(t) = \vec v_\odot + \vec v_\oplus(t) \ .
\eeq
Here $\vec v_\odot=\vec v_{\rm G}+\vec v_{\rm S}$ is the sum of the galactic rotational velocity of the local system $\vec v_\text{G}=(0, v_0, 0)$, (here $v_0\simeq 220 \pm 50$ km/s is the local DM velocity)  and the Sun's proper motion $\vec v_\text{S}\simeq  (10,13,7)$ km/s 
with absolute value $v_\odot=233\pm 50$ km/s,
while $\vec v_\oplus(t)$ is the time dependent Earth's orbital velocity with period of one year, phase around June 2$^{\rm nd}$ (when it is aligned to $\vec v_\odot$), size $v_\oplus \simeq 30$ km/s, and it is inclined of an angle $\gamma \simeq 60^\circ$ with respect to the galactic plane. More details can be found in Ref.~\cite{Fornengo:2003fm, Belli:2002yt, Savage:2008er}. Since in Eq.~\eqref{vE} the Earth's orbital velocity projected in the galactic plane $v_\oplus \cos\gamma$, is relatively small compared to $v_\odot$, we can approximate $\vec v_\text{E}(t)$ with its component directed toward the galactic center. We can then write 
\beq\label{vEarth}
v_{\rm E}(t) \simeq v_\odot+\Delta v \, \cos\left[2\pi(t-\phi)/\tau\right] \ ,
\eeq 
where $\Delta v = v_\oplus\cos\gamma\simeq 15$ km/s. Here $\phi \simeq 152.5$ days (June 2$^\text{nd}$) is the phase and $\tau\simeq 365$ days is the period of the Earth motion around the Sun.

Therefore, in general, it is expected an annual modulation in the counting rate of direct detection experiment, as the incoming flux of DM particles toward us contains a small oscillatory terms due to the Earth's proper motion around the Sun during the years. More details on this model independent signature can be found for instance in Refs.~\cite{Drukier:1986tm, Freese:1987wu}. It is worth stressing that looking for an annual modulation of the counting rate in direct detection is very challenging from the experimental point of view, simply because the modulus of the time dependent component of the velocity is relatively small compared to the constant one;  in better words, the size of the modulated signal that experiments like \DAMA\ and \CoGeNT\ are looking for is suppressed respect to the unmodulated one, due basically to the collective motion of the Solar system around the galactic center. 

\item In the non relativistic limit the differential cross section can be written in the usual form 
\beq\label{sigmaT}
\frac{\ud \sigma_{\cal N}}{\ud \ER} (v, \ER) = \frac{1}{32 \pi} \frac{1}{m_\chi^2 m_{\cal N}} \frac{1}{v^2} \overline{| \Mel_{\cal N}|^2} \ ,
\eeq
where $\overline{| \Mel_{\cal N}|^2}$ is the DM-nucleus matrix element that encodes all the information related to the nuclear physics and the nature of the interactions. Since, as already stated, the velocity of the DM particles in the vicinity of the Earth is much smaller then the speed of light, the framework of relativistic quantum field theory is not too much appropriate to study signals in this field, especially if we do not have any clue about the underlying theory that describes the DM-quarks interactions. A more useful and general framework is actually the one based on non-relativistic operators.  Indeed, since for the elastic scattering, the relevant degrees of freedom are the DM-nucleon relative velocity $\vec{v}$, the exchanged momentum $\vec{q}$, the nucleon spin $\vec{s}_N$ ($N = p, n$ can be proton or neutron) and the DM one $\vec{s}_\chi$ (if different from zero), the scattering amplitude will then be a rotationally invariant function of this variables. In this regard, with $\vec{v}$, $\vec{q}$, $\vec{s}_N$, and $\vec{s}_\chi$ \cite{Dobrescu:2006au}, we can construct a basis of $16$  operators, which includes all possible spin configurations (the complete list and the numbering of the non-relativistic operators can be found in Refs.~\cite{Fitzpatrick:2012ix, Fitzpatrick:2012ib, Anand:2013yka}). The DM-nucleon matrix element can then be written as a linear combination of these operators, with coefficients that may depend on the momenta only through the $q^2$ or $v^2$ scalars\footnote{DM models feature a long range interaction with the Standard Model fermions  provide perhaps the most notable example. Indeed, in this case, the differential cross section is enhanced at smaller exchanged momenta, due to the negative power dependence of $q$. A specific realization is offered by  models in which DM particles can carry small but nonzero electric charge, electric dipole moment  or magnetic one. Their phenomenology in the context of DM direct searches has been studied in Refs.~\cite{Pospelov:2000bq,Foot:2008nw,Foot:2010rj,Foot:2011pi,Fornengo:2011sz,Panci:2012qf,Kaplan:2011yj,Sigurdson:2004zp,Chang:2010en,Barger:2010gv,DelNobile:2012tx,Ho:2012bg} and references therein.}.   In particular it reads
\beq\label{Leff}
\Mel_N =
\sum_{i = 1}^{16} \mathfrak{c}^N_i(\lambda, m_\chi) \, \Op^\NR_i \ ,
\eeq
where the coefficients $\mathfrak{c}^N_i$ are functions of the free parameters of the underlying relativistic theory, such as mediator masses,  mixing angles, couplings constants and so on (collectively denoted by $\lambda$), and the DM mass $m_\chi$. For instance, if the interaction between a fermionic DM $\chi$ and the nucleon $N$ is described by the ``standard'' SI Lagrangian $g_N/\Lambda^2 \ \bar \chi \chi \, \bar{N} N$, the only non-relativistic operator involved will be the identity ($\Op^\NR_1=\unop$),  Before moving on, it is worth stressing that the operators above, which entirely describe the non-relativistic physics of the DM-nucleus scattering, are many and they can either depends on the exchanged momenta $\vec q$ or the relative velocity $\vec v$. As a consequence, focussing in just one operator (e.g. $\Op^\NR_1$) is not the most model independent way to study signals in direct DM searches, since  theories can predicted several operators entering together with the possibility that some of them may even interfere. 

\begin{figure}[t]
\centering
\includegraphics[width=.475\textwidth]{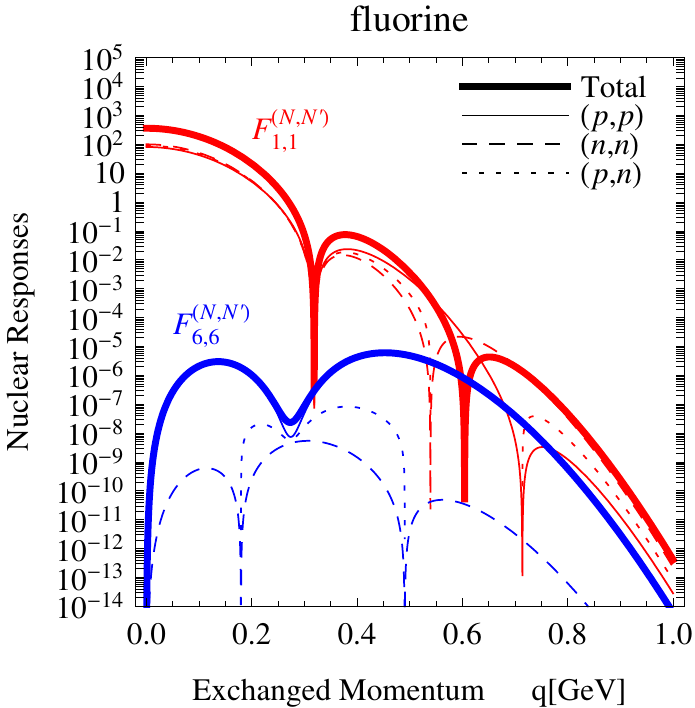}\quad
\includegraphics[width=.475\textwidth]{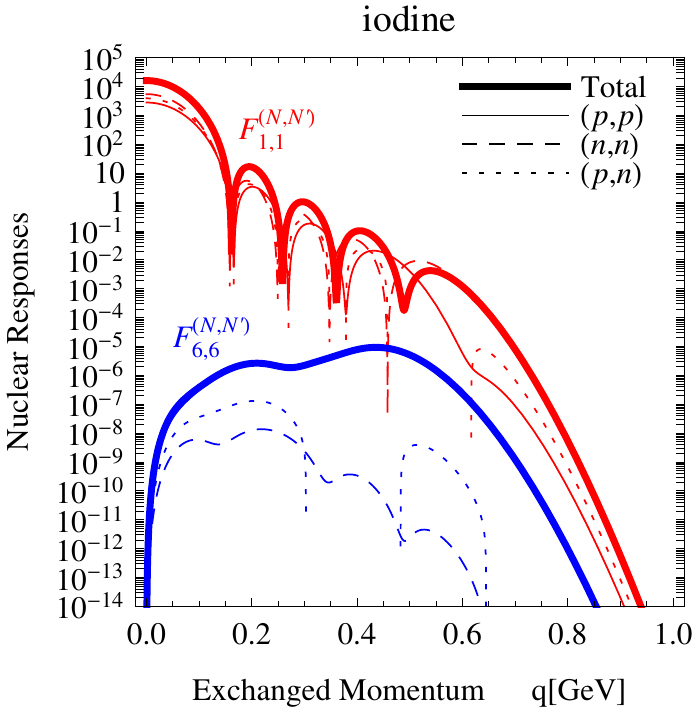}
\caption{\em \small \label{fig:1} 
 {\bfseries Nuclear Responses} of the fluorine (left-hand side) and the iodine (right-hand side) targets considering two completely different types of interactions. On a more specific level, we show, in red the nuclear responses for the ``standard'' SI contact interaction, while in blue those for another kind of interaction described in the non-relativistic limit by the operator $\Op_6^\NR=(\vec s_\chi\cdot\vec q)(\vec s_N \cdot\vec q)$. The different hatching refers instead to the possible choices of nucleon pairs in the nucleus. In particular, the solid, dashed and dotted lines are for $(p,p)$, $(n,n)$ and $(p,n)$ pairs, while the thick ones are obtained by summing over all of them (total nuclear responses). Notice that the total nuclear responses for the ``standard'' SI contact interaction (thick red lines) reduce to $A_{\cal N}^2 F_{\rm Helm}^2(q)$, where $F_{\rm Helm}^2(q)$ is the customary Helm form factor \cite{HelmFF}.}
\end{figure}

Since now the nucleus is of course made of protons and neutrons, one has to correct the DM-nucleon matrix element \eqref{Leff} with the nuclear responses, which are a sort of form factors that take account of the finite size of the target nuclei. According to Eq.~(55) of \cite{Fitzpatrick:2012ix} we can then write the spin-averaged amplitude squared for scattering off a target nucleus $\cal N$ with mass $m_{\cal N}$ as
\beq\label{Mel}
\overline{\left| \Mel_{\cal N} \right|^2} = \frac{m_{\cal N}^2}{m_N^2} \sum_{i, j = 1}^{16} \sum_{N, N' = p, n} \mathfrak{c}^N_i \mathfrak{c}^{N'}_j F_{i, j}^{(N, N')} \ .
\eeq
The $F_{i, j}^{(N, N')}(v, \ER, \cal N)$ are the nuclear responses  and they depend critically on the type of scattering nucleus $\cal N$: they are also functions of $m_\chi$, $v$ and the nuclear recoil energy $\ER = q^2 / 2 m_{\cal N}$. In Fig.~\ref{fig:1}, the nuclear responses for both light element (on the left panel) and heavy one (on the right panel), considering two completely different kind of interactions, are depicted. On a more specific level, in thick red the total nuclear response for the standard SI interaction is shown. As we can see, in the energy range of interest in direct DM experiments (few keV), it is enhanced by the canonical $A_{\mathcal N}^2$ factor: this is due to the fact that, in this case, the incoming DM particles see the nucleus as a point-like object with $A_{\mathcal N}$ scatter centers. On the other hand, we show in blue the nuclear responses of a completely different type of interaction which is both momenta and spin dependent ($\Op_6^\NR=(\vec s_\chi\cdot\vec q)(\vec s_N \cdot\vec q)$). This type of interaction can give rise if the DM-nucleus interaction is mediated by a pseudo-scalar particle. We can see that the nuclear responses are dominated by the scattering with protons, simply because neither the fluorine nor the iodine have unpaired neutrons and moreover the interaction goes to zero in the small momenta transferred limit. 

A complete set of these nuclear responses for each pairs of nucleons $(N,N')$, each pairs of non-relativistic operators $(i,j)$ and for several target nuclei $\cal N$, has been provided in the appendices of Ref.~\cite{Fitzpatrick:2012ix}. This is extremely useful because,  for the first time,  different types of interactions can be studied in a more general ground. However, since this kind of computations are quite new, the uncertainties, especially for the spin, momenta and velocity  dependent interactions, are still quite large. 

\end{itemize}

Having a disposal the general relation of the DM-nucleus matrix element \eqref{Mel}, the differential cross section \eqref{sigmaT} and in turn the rate of nuclear recoil \eqref{RN} can be re-written, following Ref.~\cite{DelNobile:2013sia}, in a very general way. It reads

\beq\label{Rate}
\frac{\ud R_{\cal N}}{\ud \ER} =
X \, \xi_{\cal N} \sum_{i, j = 1}^{16} \sum_{N, N' = p, n} \mathfrak{c}^N_i(\lambda, m_\chi) \, \mathfrak{c}^{N'}_j(\lambda, m_\chi) \, \mathcal{F}_{i, j}^{(N, N')}(\ER, {\cal N}) \ ,
\eeq
where we defined
\beq
X \equiv. \frac{\rho_\odot}{m_\chi} \frac{1}{32 \pi} \frac{1}{m_\chi^2 m_N^2} \ .
\eeq
and
\beq
\mathcal{F}_{i, j}^{(N, N')}(\ER, {\cal N}) \equiv \int_{\vmin(\ER)}^{\vesc} \hspace{-.50cm} \ud^3 v \, \frac{1}{v} \, f_\text{G}(\vec v+ \vec v_{\rm E}(t)) \, F_{i, j}^{(N, N')}(v, \ER, {\cal N}) \ .
\eeq

\subsubsection{``Standard'' SI contact interaction}\label{UsualSI}
To give  a concrete example, in the following the coefficient and the nuclear responses for the ``standard'' SI 
contact interaction are  showed explicitly. In particular, when experiments present results in terms of this interaction, they implicitly assume (probably inspired by supersymmetric neutralino scattering) the following  DM-nucleon effective Lagrangian 
\begin{align}
\label{LagSI}
\Lag_{\rm SI}^N &= \lambda_{\rm SI}^N \, \bar \chi \chi \bar N N \ .
\end{align}
From this the DM-nucleon matrix element can be obtained by contracting Eq.~\eqref{LagSI} 
with initial and final states of the scattering. Performing then the non-relativistic limit one gets:
\begin{align}
\label{MelSI}
\Mel_{\rm SI}^N & \equiv  \langle \chi N |  \Lag_{\rm SI}^N | \chi N \rangle  \simeq 4\, \lambda_{\rm SI}^N\, m_\chi m_N \,\Op_1^\NR \ ,
\end{align}
where $\Op_1^\NR =\unop$ 
is the operator that describes the non-relativistic limit of Eq.~\eqref{LagSI} 
and the coefficient is obviously given by $\mathfrak{c}^N_1(\lambda_{\rm SI}^N, m_\chi) = 4\, \lambda_{\rm SI}^N \, m_\chi m_N $.

For the SI scattering, DM-proton and DM-neutron couplings are customarily assumed to be equal and thus $\lambda_{\rm SI}^p=\lambda_{\rm SI}^n\equiv \lambda_{\rm SI}$ (isospin violating process are not taken into account). Plugging  back then the coefficient in the general equation \eqref{Rate}, and defining $\sigma_{\rm SI}=\lambda_{\rm SI}^2/\pi \cdot \mu_N^2$ ($\mu_N$ is the DM-nucleon reduced mass) to make contact with the usual physical cross sections, the differential rate of nuclear recoil reduces to the very well know compact form 
\beq\label{RateSI}
\frac{\ud R_{\cal N}}{\ud \ER} =\frac{\xi_{\cal N}}{m_{\cal N}} \frac{\rho_\odot}{m_\chi}  \left[ \frac{m_{\cal N}}{2 \mu_N^2} \sigma_{\rm SI} \, F^{\rm tot}_{\cal N}(\ER)\,  \mathcal I (\ER) \right]\ ,
\eeq
where we define the total SI  nuclear response as
\beq
F^{\rm tot}_{\cal N}(\ER)=\sum_{N, N' = p, n} \hspace{-.25cm} F_{1,1}^{(N, N')}(\ER, {\cal N}) \ .
\eeq
Here, since $\lambda_{\rm SI}^p=\lambda_{\rm SI}^n$, the total nuclear response reduces to the square of the customary Helm form factor \cite{HelmFF} multiplied by the coherent enhancement of the cross section $A_{\cal N}^2$. In particular it reads $F^{\rm tot}_{\cal N}(\ER) = A_{\cal N}^2 F_{\rm Helm}^2(\ER)$. The function $ \mathcal I(\ER) = \int_{v_{\rm min}}^{v_{\rm esc}} \hspace{-.05cm}\ud^3 v f_\text{G}(\vec v+\vec v_{\rm E}(t)) / v$ is instead the velocity integral encountered many times in the literature.  It is worth noticing that $ \mathcal I(\ER)$ can be written in this way because in the ``standard'' SI interaction the operator that the describes the non-relativistic limit of the effective Lagrangian does not carry any dependences on the relative velocity $\vec v$. In Sec.~\ref{BoringSI} the  interpretation of the experimental results in terms of the ``standard'' SI interaction is briefly reviewed.

\subsection{Experimental Observables}\label{ExpObs}
Since, as already stated, the Earth's orbital velocity projected in the galactic plane is relatively small compare to the drift velocity of the Sun, we can then expand the recoil rate \eqref{Rate}, assuming that the velocity distribution is not strongly anisotropic. Then, by means then of the Eq.~\eqref{vEarth}, one gets 
\beq
\frac{\ud R_{\cal N}}{\ud \ER}(t)\simeq \left.\frac{\ud R_{\cal N}}{\ud \ER} \right|_{v_{\rm E}=v_\odot}+\frac{\partial}{\partial v_{\rm E}}\left.\frac{\ud R_{\cal N}}{\ud \ER} \right|_{v_{\rm E}=v_\odot} \hspace{-.25cm}\Delta v \cos\left[2\pi(t-\phi)/\tau\right] \ .
\eeq

In order now to properly reproduce the experimental  rate of nuclear recoil and in turn the expected number of events in a certain energy bin, one should take into account the response of the detectors as well. It can be done by the following energy convolution and transformation 
\beq
\label{dRdEd}
\frac{\ud R}{\ud \Ed} = \sum_{\cal N} \epsilon(\Ed) \int_0^\infty \ud \ER \, \mathcal{K}_{\cal N}(q_{\cal N} \ER, \Ed) \, \frac{\ud R_{\cal N}}{\ud \ER}\left(\ER \right) \ ,
\eeq
where $E'$ is the detected energy, the functions $\mathcal{K}_{\cal N}(q_{\cal N} \ER, \Ed)$ and  $\epsilon(\Ed)$ are the energy resolution centered in $q_{\cal N} \ER$ and the detector's efficiency respectively. Here the sum runs over the different species of the detectors (e.g.~\DAMA\ is a multiple target experiment composed by crystal of sodium and iodine) and $q_{\cal N}$ is the so-called quenching factor that takes account of the partial recollection of the released energy in the detector. After convolving with all the experimental effects, Eq.~\eqref{dRdEd} must be  averaged over the energy bins of the detector. For each energy bin $k$, we then obtain the number of the unmodulated events $N^{\rm th}_{0\,k}$ and of the modulated ones $N^{\rm th}_{\text{m}\,k}$, as
\beq\label{ExpTotalRateTay1}
N^{\rm th}_{0 \, k}=w_k \int_{\Delta E_k}\ud \Ed\,\left.\frac{\ud R}{\ud \Ed}\right|_{v_{\rm E}=v_\odot} \ ,
\eeq
\beq\label{ExpTotalRateTay2}
N^{\rm th}_{\text{m} \, k}=w_k \Delta v \int_{\Delta E_k}\ud\Ed\,
\left.
\frac\partial{\partial v_{\rm E}}\frac{\ud R}{\ud\Ed}
\right|_{v_{\rm E}=v_\odot} \ .
\eeq
where $w_k$ is the exposure (expressed kilograms per days) and $\Delta E_k$ is the width of the $k^{\rm th}$-esime energy bin.   $N^{\rm th}_{0 \, k}$ and $N^{\rm th}_{\text{m} \, k}$ are the relevant quantities that one can use for the analysis of the experiments which address the annual modulation effect, namely \DAMA\ and \CoGeNT. For the other experiments, only $N^{\rm th}_{0 \, k}$ is instead relevant.

Collecting all the elements in the previous equations we expand Eqs.~(\ref{ExpTotalRateTay1}, \ref{ExpTotalRateTay2}) and write

\beq\label{Nkth0}
N^{\rm th}_{0 \, k} = X \sum_{i, j = 1}^{16} \sum_{N, N' = p, n} \mathfrak{c}^N_i(\lambda, m_\chi) \, \mathfrak{c}^{N'}_j(\lambda, m_\chi) \, \left.\tilde{\mathcal{F}}_{i, j}^{(N,N')}(m_\chi, k)\right|_{v_{\rm E}=v_\odot} \ ,
\eeq
\beq\label{Nkthm}
N^{\rm th}_{\text{m} \, k} = X \Delta v \sum_{i, j = 1}^{16} \sum_{N, N' = p, n} \mathfrak{c}^N_i(\lambda, m_\chi) \, \mathfrak{c}^{N'}_j(\lambda, m_\chi) \, \frac\partial{\partial v_{\rm E}}\left.\tilde{\mathcal{F}}_{i, j}^{(N,N')}(m_\chi, k)\right|_{v_{\rm E}=v_\odot}  \ .
\eeq

Here $\tilde{\mathcal{F}}_{i, j}^{(N,N')}(m_\chi, k)$ is a sort of {\em integrated form factor} that encapsulates all the information related to nuclear physics, astrophysics and the detector dependency of the rate. There is one of these factors for each energy bin $k$ of each experiment under consideration, and for each choice of the operators pair $(i,j)$ and nucleons pair $(N,N')$. It reads explicitly
\beq\label{integratedFF}
\tilde{\mathcal{F}}_{i, j}^{(N,N')}(m_\chi, k) = w_k \sum_{\cal N} \xi_{\cal N}
\int_{\Delta E_k} \hspace{-.10cm} \ud \Ed \, \epsilon(\Ed) \int_0^\infty \ud \ER \, \mathcal{K}_T(q_{\cal N} \ER, \Ed) \, \mathcal{F}_{i, j}^{(N, N')}(\ER, \cal N) \ .
\eeq

It is worth noticing  that the linearity of the expected number of events in the \emph{integrated form factors} is a fundamental ingredient. This, indeed, lets us parametrize the model dependent part of the rate from the model independent one, encapsulated in the function $\tilde{\mathcal{F}}_{i, j}^{(N,N')}(m_\chi, k)$. Therefore, if the experimental collaborations, instead of  presenting  the experimental results in terms of a specific DM model, release all of this finite number of \emph{integrated form factors}, one will be easily able to obtain the expected number of events for any kind of interactions, whose particle physics is completely incapsulated in the coefficient $\mathfrak{c}^N_i$. A first example of the potentiality of such parametrization has been presented in Ref.~\cite{DelNobile:2013sia}. In particular the authors provide a self-contained set of numerical tools based on the \emph{integrated form factors}  above,  to derive the bounds from some current direct detection experiments on virtually any arbitrary model of DM elastically scattering on target nuclei.

\section{Direct detection status}\label{DDStatus}

\subsection{Experimental Landscape}\label{ExpStatDirectD}
The elusive nature of DM particles makes their detection a challenge for experimentalist: in fact, considering typical atomic masses for the target nuclei of $m_{\cal N}=100$ GeV, and typical properties for the DM halo (Maxwell-Boltzmann distribution with dispersion velocity $v_0=220$ km/s), DM particles should induce tiny nuclear recoil in the range $1\div 10^3$ keV with a total rate lower then 1 cpd/kg/keV\footnote{Here cpd refers to counts per days.}. Due to this rare phenomena,  the experimental priorities in this field are:
\begin{itemize}
\item[$\diamond$] the detectors must work deeply underground in order  to avoid the high rate interaction induced by cosmic rays scatter off on target nuclei.  
\item[$\diamond$]  they must use active shields and very clean materials against the residual radioactivity in the tunnel (mostly $\alpha$-particles, neutrons and photons).
\item[$\diamond$]  they must distinguish multiple scattering, simply because DM particles do not interact twice in the detector, being of course weakly interacting particles.

\end{itemize}

A variety of different experimental techniques with the aim of measuring the tiny energy released by a DM interaction have been developed. In the following the main channels by which the scattered nucleus can deposit energy in the detectors are summarized:

\begin{itemize}
\item \emph{Scintillation detection:} a particle interacting within a scintillating target induces the emission of light produced by the de-excitation of exited atoms. This light can be detected by appropriate photomultipliers. Typically NaI(Tl) and xenon are used as scintillators.  
\item \emph{Ionization detection:} a particle interacting inside a target produces an amount of free electron-ion pairs that can be detected with a collecting drift field and a device sensitive to the electric charge.  
\item \emph{Phonon detection:} a particle interacting inside a detectors releases a tiny energy deposition with a subsequently increase of the temperature.  Cryogenic apparatus working at very low temperature (around few mK) are able to measure this small variation making this detection technique possible. %
\end{itemize}
 
Since various types of interacting particles release a different amount of energy in the channels commented upon above, for a better rejection of the background events, most of the experiments are designed to be sensitive to more than one of them. Indeed, thanks to this ability, the ratio between two channels can be used to distinguish between nuclear (due to a DM interaction) and electromagnetic recoils.  

The  only  two experiments that do not use this kind of experimental technique   are  \DAMA\ and \CoGeNT, that, perhaps not coincidentally, were the only two detectors (until the \CRESST\ and recently the \CDMSSi\ results) observing an excess in their counting rate.  In particular, since both the experiments are not able to disentangle the nuclear recoil signals from the electromagnetic ones, they infer the presence of DM in our halo by exploiting the model independent annual modulation signature in the counting rate. As follows, I review the most important experiments in this field.

\subsubsection{DAMA} 
The \DAMA\ experiment, located at the National Laboratory of Gran Sasso, is an observatory for rare processes made of highly radiopure scintillators (NaI(Tl) crystals). In particular the former \DAMANaI\ and the current \DAMALibra\ experiments have the main aim of investigating the presence of DM particles in the galactic halo by exploiting the model independent annual modulation induced by a DM interaction. 

The \DAMA\ detector is only able to measure  the fraction of energy deposited in scintillation light, while the phonon excitation due to multiple nuclear interactions are not detected. This effect, as commented upon in Sec.~\ref{ExpObs}, is taken into account by the so-called quenching factor. In particular, for NaI(Tl) crystals, one normally considers $q_{\rm Na}\simeq 0.3$ ad $q_{\rm I}\simeq 0.09$ \cite{Bernabei:1996vj}.  It is, however, known \cite{Bernabei:1996vj, Bernabei:2007hw, Drobyshevski:2007zj} that some scattered nuclei can travel long distances along the crystallographic axes and planes   without colliding with other nuclei. This process is called channeling effect, and since no phonon excitation are produced, the scattered nuclei deposit all the energy in the detector electromagnetically ($q_{\mathcal N}\simeq 1$).  The fraction of channeling depends on the nucleus itself, on the recoil energy and strongly on the temperature. It has been calculated for instance in Ref.~\cite{Bernabei:2007hw} and, in particular, it has been found that, for low energy recoiling Na and I ions, the fraction of channeled events can be relatively large. Nevertheless, this result has not been confirmed by other theoretical  \cite{Graichen:2002kg,Feldstein:2009np,Bozorgnia:2010xy} and experimental studies \cite{Collar:2013gu}, which instead suggest that the channeling effect in NaI crystals is negligible.


The \DAMA\ collaboration published results of the combined \DAMANaI\ and \DAMALibra\ experiments \cite{Bernabei:2008yi,Bernabei:2010mq}, corresponding to an exposure of 1.17 ton$\cdot$yr for a target.  They observed a cosine-like modulation, present only in the \emph{single-hit} events, with a measured period $\tau=(0.999\pm 0.002$) yr and a phase $t'=(146\pm7)$ days well compatible with the roughly 152.30 days expected for DM signals (see previous section). The modulation is present only in the low energy window ($2-6$) keVee\footnote{Here, keVee refers to  electron equivalent recoil energy. This must be converted with the quenching factor in order to get the total energy released to the nucleus by the scattering process ($\ER$[keV] $\equiv \ER$[keVee]$/q_{\mathcal N}$). }  and its amplitude $N^{\rm exp}_{\text{m}\,(2-6)}$ is equal to ($0.0116\pm0.0013$) cpd/(kg$\cdot$keV) at 8.9 $\sigma$ CL \cite{Bernabei:2010mq}. 


\subsubsection{CoGeNT} 
The \CoGeNT\ experiment employs p-type point-contact Germanium detectors operating in the Soudan Underground Laboratory. 

Like \DAMA, the experiment performs just one of the techniques commented upon above, namely the p-type point-contact germanium detectors only measure the fraction of energy deposited by incident particles in the form of ionization. The lack of energy measured is again taken into account by the quenching factor, that can be extracted through the following empirical relation $E_{\rm det} = 0.2 \, \ER^{1.12}$ \cite{Barbeau:2007qi}. By virtue of its low energy threshold (0.4 keVee) and the ability to reject surface backgrounds, this type of detector is particularly sensitive to light DM candidates, although large background contamination may be present for these low energies. 

In 2010 the collaboration have reported a step rise of nuclear recoil spectrum at low energy which is not directly explainable in terms of known radioactive background \cite{Aalseth:2010vx}. The energy region probed by \CoGeNT\ partially overlap with the one in which the \DAMA\ apparatus reported the annual modulation signal, and therefore it is natural to interpret the excess at low energy due to a DM interaction. In view of that, fifteen months of cumulative \CoGeNT\ data (442 live days) have been also examined in order to look for an annual modulation signature \cite{Aalseth:2011wp}. In particular  the \CoGeNT\ data seem to favor a cosine-like modulation slightly shifted in phase respect to the one measured by \DAMA. The modulation is again only present at low energy with a statistical significance of 2.8 $\sigma$ CL, limited by the short exposure.

Very recently, the collaboration presented a 3.4 years data taking, with an  improved analysis that allows a better discrimination between pure bulk  and pure surface events \cite{Aalseth:2014eft,Aalseth:2014jpa}. In the same energy region, where in 2010 they observed a step rise of the nuclear recoil spectrum, they still reported a  preference for an annual modulation in the pure bulk counting rate. Although the statistical significance of the modulation is modest (2.2 $\sigma$ CL), the phase is  compatible with the one observed by \DAMA.
%
%
%
%
%

\subsubsection{CRESST-II}
The \CRESST\ cryogenic Dark Matter experiment, located at the national laboratory of Gran Sasso, employ 300 g of scintillating CaWO$_4$ crystals. 

In particular the eight modulus of the experiments measure the deposited energy in the form of phonons and scintillation light. The former provides a precise measure of the energy deposited, while the ratio between them gives an excellent rejection of the background events.  

In 2011 the collaboration completed a 730 kg$\cdot$days  data taking  \cite{Angloher:2011uu}. In particular, 67 events were found in the DM acceptance region with an expected background contributions in the same band which is not sufficient to explain all the observed events. The resulting statistical significances, at which  the background-only hypothesis is rejected, 
is roughly 4.5 $\sigma$ CL.

\subsubsection{CDMS} 
The \CDMS\ experiment, located at the Soudan Underground Laboratory, is composed by  Germanium and Silicon cryogenic detectors. 

In particular $19$ Germanium and $11$ Silicon detectors measure the  deposited energy in the form of phonons and ionization through superconducting technology. Thanks to the ability of detecting these two signals, the rejection of the electromagnetic recoils can be obtained by the ratio among them.  

In 2009 the \CDMS\ collaboration \cite{Ahmed:2009zw} has reported no significant evidence for DM interaction. In particular, only Germanium detector were used with an exposure of 612 kg$\cdot$days. The collaboration observed two events in the signal region with energy 12.3 keVnr and 15.5 keVnr\footnote{Here, keVnr refers to  nuclear recoil energy. Unlike the electron equivalent  recoil energy, it already represents the total energy released to the nucleus by the scattering process ($\ER$[keV] $\equiv \ER$[keVnr]).}, against an expected background of radiogenic neutrons, cosmogenic neutrons and misidentified surface events equal to roughly 0.9 events. 

More recently, the collaboration has also presented results based on the Silicon analysis~\cite{Agnese:2013cvt, Agnese:2013rvf}. In Ref.~\cite{Agnese:2013rvf}, a blind analysis of 140.2 kg$\cdot$days revealed three DM candidate events in the signal region with an expected background of roughly 0.62 events. A profile likelihood ratio statistical test of the three events yields a 0.19\% probability when the known-background-only hypothesis is tested, against the alternative DM+background hypothesis.

\subsubsection{Xenon Experiments} 

Direct detection experiments based on liquid/gaseous xenon have done stunning progresses in the last decade. Indeed,  thanks to the advantage in scale-ability and to the good reconstruction of the three-dimensional coordinates, these experiments are among the largest in terms of fiducial mass. 

Discrimination between nuclear and electromagnetic recoils is achieved by the ratio between the scintillation signal due to a particle interaction in the liquid xenon, and the subsequent ionization signal in the gas phase of the detector. Furthermore, the large mass number of the xenon nuclei makes them an excellent target for the detection of DM particles with SI interaction. However, the experiments based on liquid xenon are also sensitive to the $n$-DM spin dependent interaction, by virtue of the unpaired neutron of the $^{129}$Xe and $^{131}$Xe isotopes.

In the following the results of the main experiments using the double-phase xenon technology are summarized:

\begin{itemize}

\item[$\diamond$] \emph{XENON experiment:} The \XENON\ detector is a two-phase time projection chamber located at the national laboratory of Gran Sasso.  
The first stage of the experiment was installed underground  during March 2006, and it searched for DM interactions until October 2007. In 2008 the collaboration reported a blind-analysis  with an exposure of $ 5.4 \times  58.6$ kg$\, \cdot \,$days, that yields no significance evidence for DM interactions \cite{Angle:2007uj}. 
  
More recently the \XENONten\ experiment has been superseded by  \XENONhundred\ with more than one order of magnitude improvement in sensitivity. In Ref.~\cite{Aprile:2012nq}, the collaboration reported the results of the last run, a blind analysis with an exposure of $ 34 \times 224.6$ kg$\, \cdot \,$days which again yielded no evidence for DM interactions. In particular, the two observed events in the pre-defined nuclear recoil energy range of $6.6 - 43.3$ keVnr were consistent with the background expectation of roughly 1 event.

\item[$\diamond$] \emph{LUX experiment:} The \LUX\ experiment is a dual-phase xenon time-projection chamber operated at the Sanford Underground Research Facility in South Dakota. Like the \XENONhundred\ experiment, it measures both the scintillation light and the ionization in the gas phase to disentangle the nuclear recoils from the electromagnetic ones. In Ref.~\cite{Akerib:2013tjd}, a non-blind analysis was performed on data, collected with an exposure of $118.3 \times 85.3$ kg$\, \cdot \,$days. After  cuts, 160 events were found within the signal energy region. The collaboration found that all of these events are compatible with the expected electron recoil background.
\end{itemize}

\subsubsection{COUPP}  

The \COUPP\ experiment operating at SNOLAB in Ontario is a 4 kg CF$_3$I  bubble chamber. By virtue of its unpaired proton, the fluorine nucleus gives an excellent sensitivity for $p$-DM spin-dependent interactions, while the iodine enhances the sensitivity for the SI ones.

Particles entering the liquid in the superheated phase create a ionization tracks, around which the liquid vaporizes, forming microscopic bubbles. Bubbles grow in size as the chamber expands, making the detector able to record them both photographically and by pressure and acoustic measurements. The discrimination between nuclear and electromagnetic recoils can be achieved by choosing an appropriate chamber pressure and temperature. Indeed, under this condition, the  abundant gamma-ray and beta-decay backgrounds do not nucleate bubbles. 

In Ref.~\cite{Behnke:2012ys}, data obtained for an effective exposure to single recoil events of 437.4 kg$\, \cdot \,$days (taking into account the 79.1\% detection efficiency) were presented. Twenty single nuclear recoil events passing all the analysis cuts were observed over the three energy bins.  Due to uncertainties in the neutron background estimation, the collaboration has not attempted any background subtraction and instead has treated all of them as DM events. 

\begin{figure}[t]
\centering
\includegraphics[width=.57\textwidth]{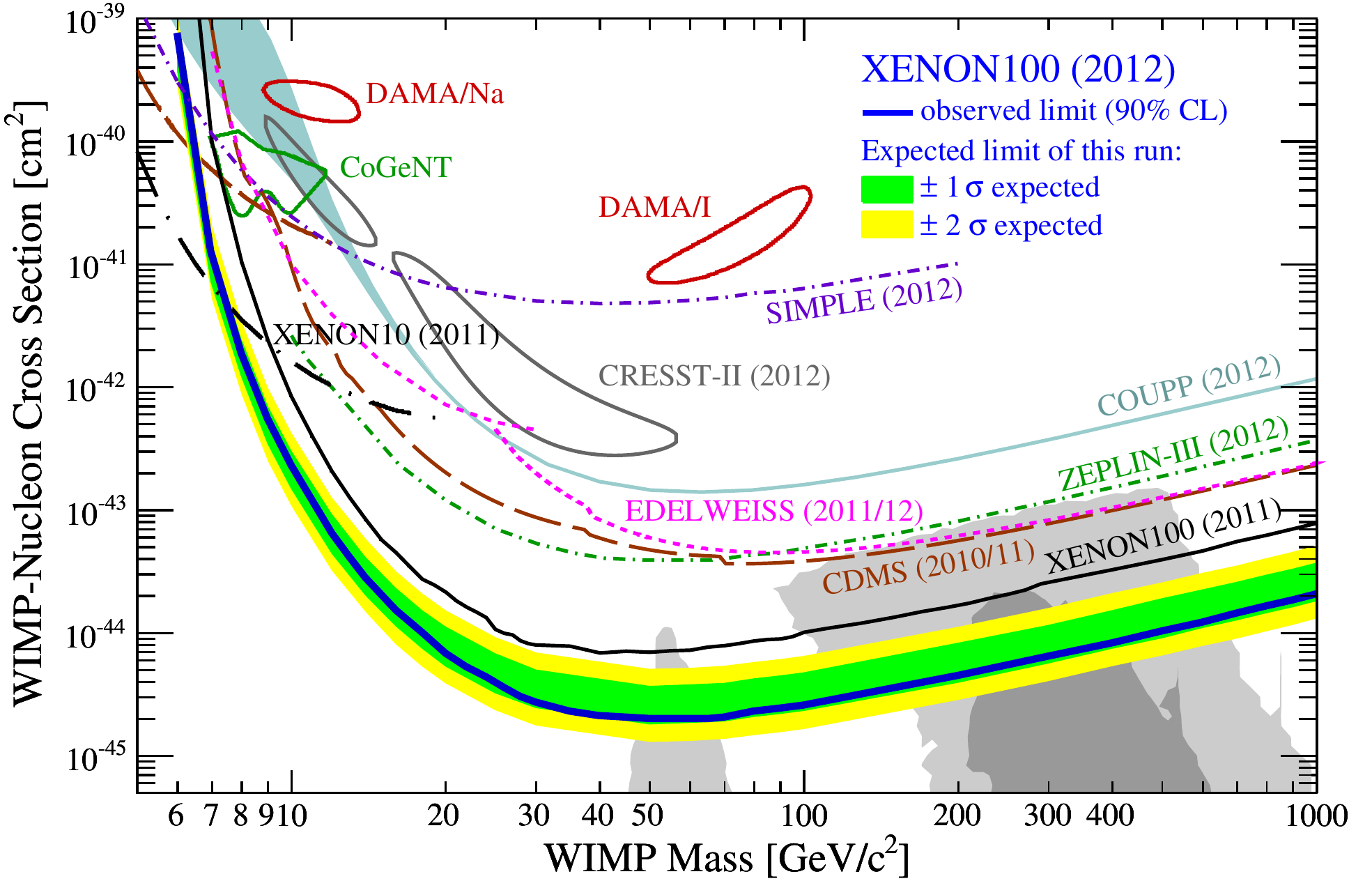}\quad
\includegraphics[width=.38\textwidth]{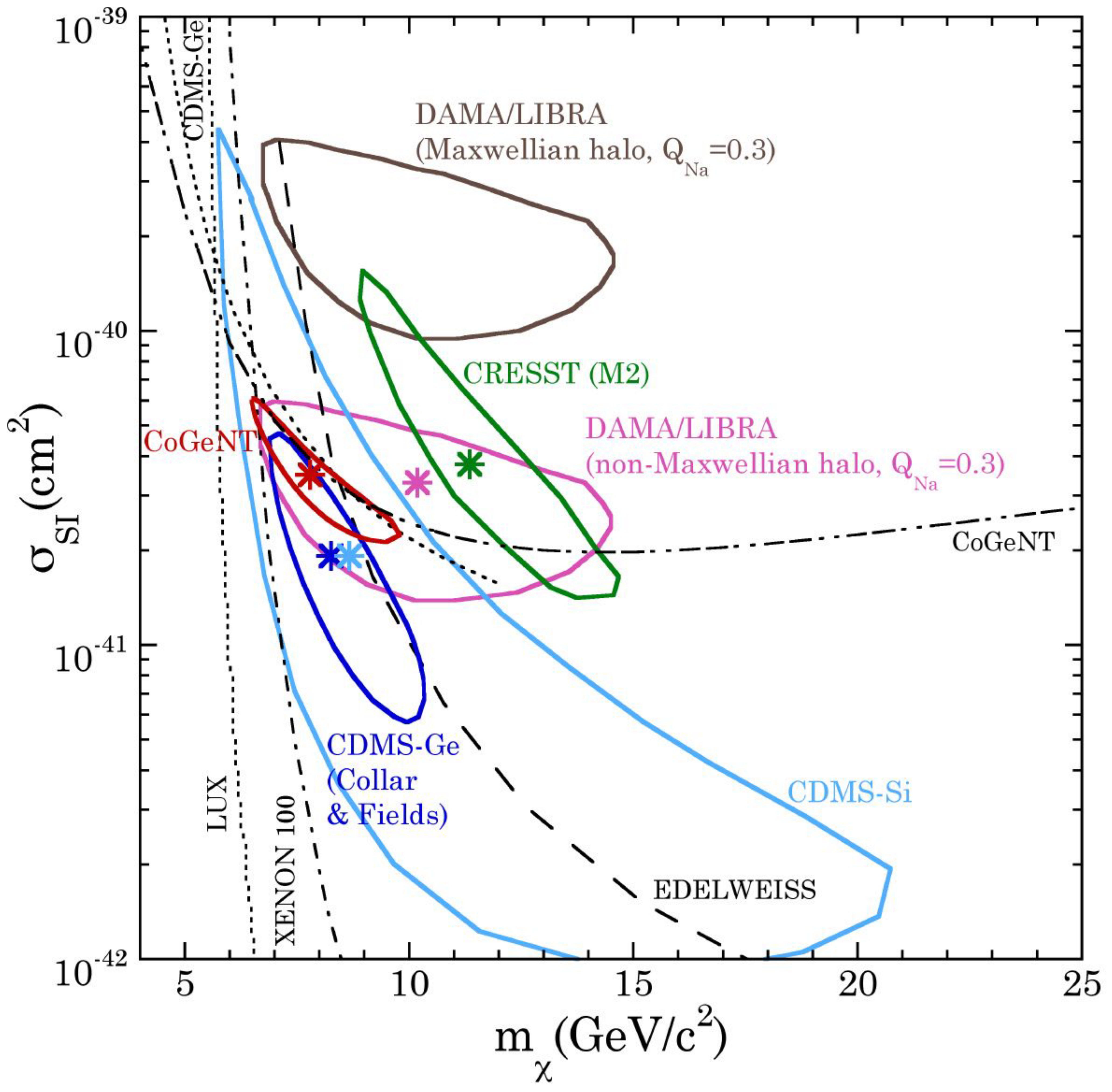} 
\includegraphics[width=.51\textwidth]{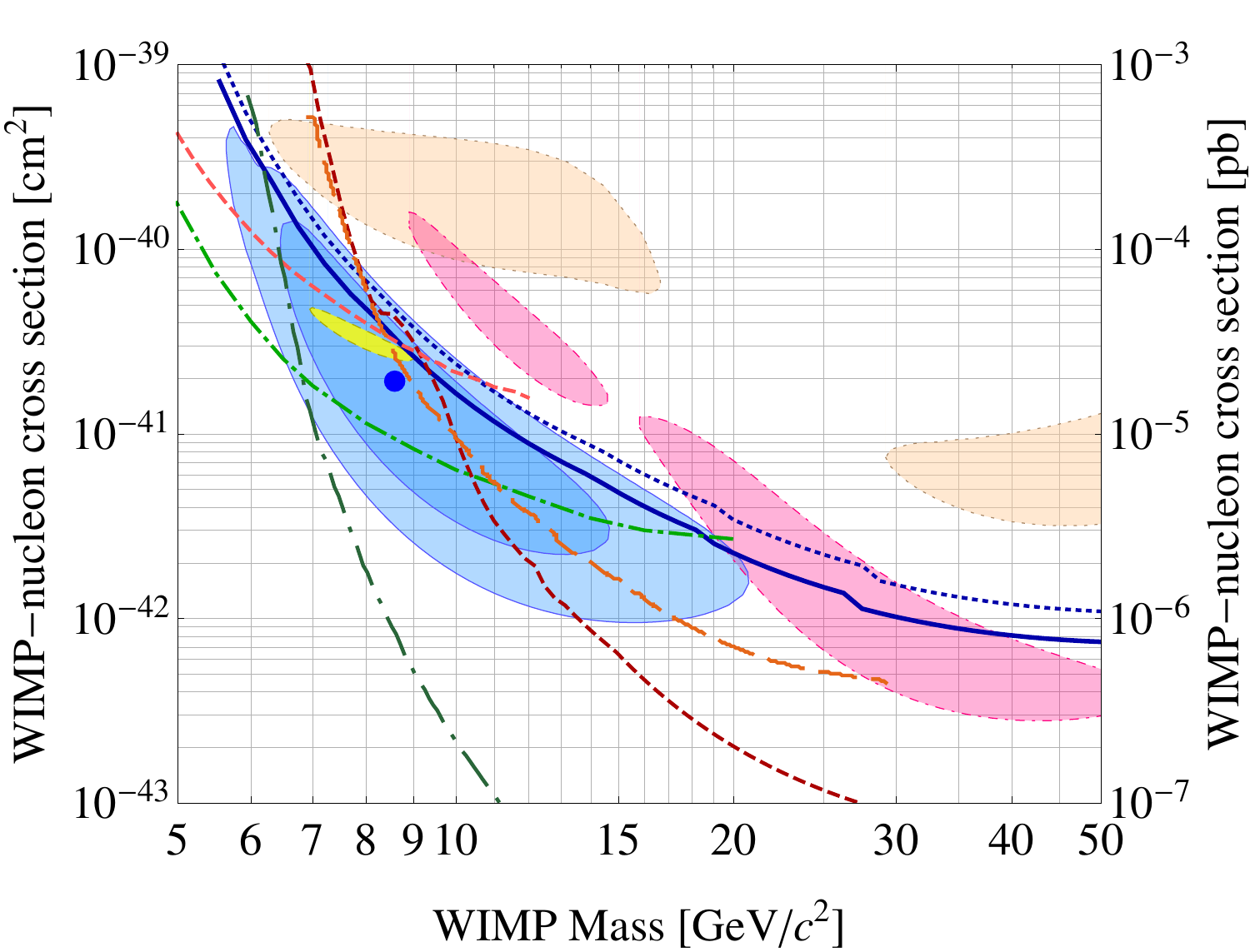}\quad
\includegraphics[width=.44\textwidth]{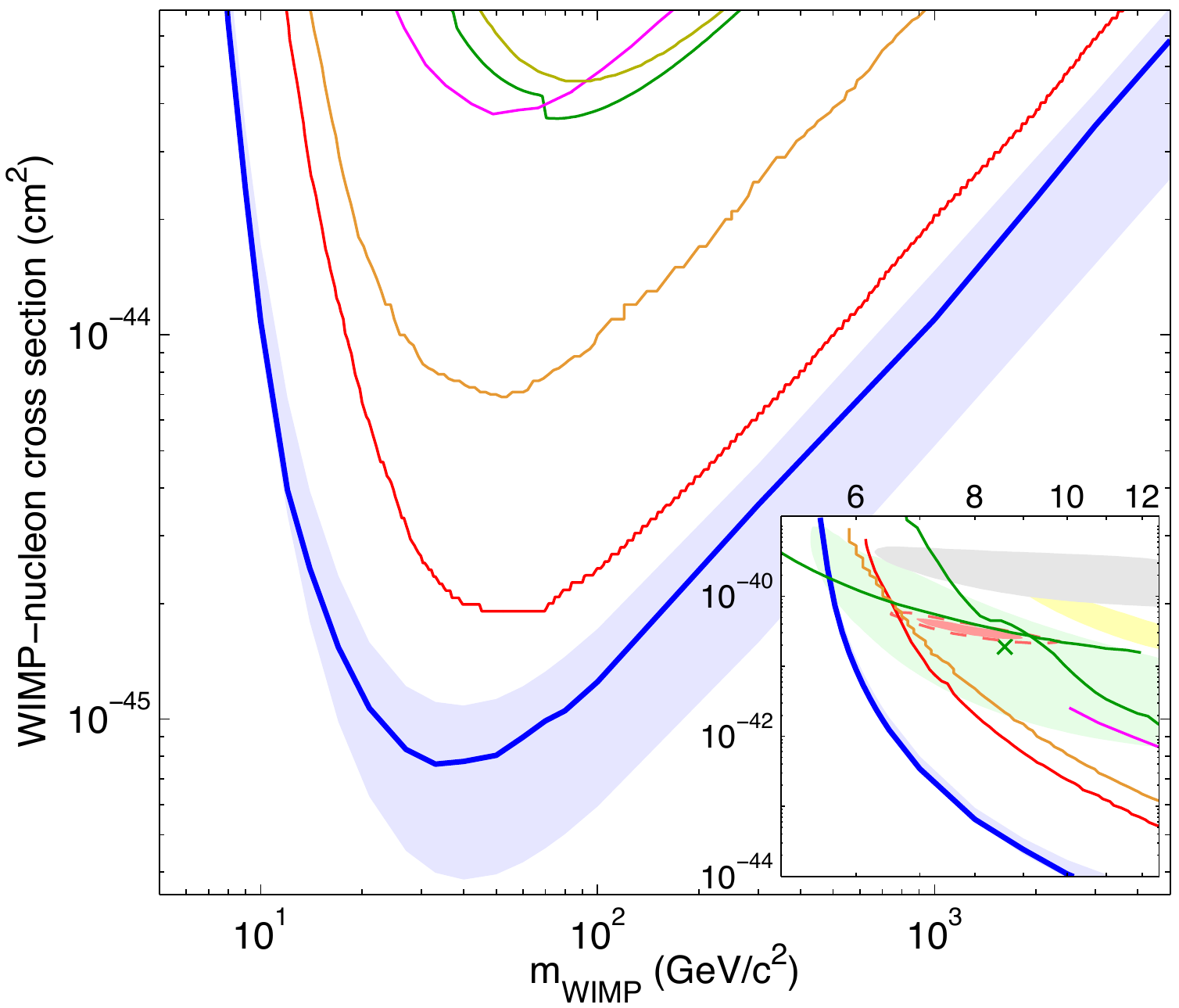} 
\caption{\em \small \label{fig:2} 
{\bfseries A compilation of the allowed regions and the constraints} for the ``standard'' SI contact interaction in the ($m_\chi,\sigma_{\rm SI}$) plane. The plots on the top row are taken from \cite{Aprile:2012nq} (left) and from \cite{Aalseth:2014eft} (right), while those in the bottom row are taken from \cite{Agnese:2013rvf} (left) and from \cite{Akerib:2013tjd} (right). The galactic halo has been assumed in the form of an isothermal sphere with velocity dispersion $v_0=220$ km/s and local DM energy density $\rho_\odot=0.3$ GeV/cm$^3$.}
\end{figure}

\subsection{Interpretation in terms of the ``standard'' SI interaction}\label{BoringSI}
There are many studies in literature that try to analyze the implications of the results of \DAMA\, the anomalies in \CoGeNT, \CRESST\ and recently in \CDMSSi\, together with the null results from other experiments in terms of specific DM models and interactions. A routine way to compare results from different  DM  experiments is by assuming a particular DM velocity distribution and a certain type of DM-nucleus interaction. The customary choices that can  often be found in literature are: $i$) a truncated Maxwell-Boltzmann velocity distribution with isotropic velocity dispersion $v_0=220$ km/s; $ii$) DM particles coupling through a contact SI interaction with equal strength to the protons and neutrons. In this case one customarily chooses the total DM-nucleon cross section $\sigma_{\rm SI}$ defined in Sec.~\ref{UsualSI} together with the DM mass $m_\chi$ as free parameters, since the bounds and the allowed regions from different experiments can be compared on the same plot.

\medskip
In Fig.~\ref{fig:2} a compilation of the allowed regions of the positive results experiments and the constraints coming from null results taken from different experimental collaborations are shown. Without entering in the details of the different analysis, we summarize the interpretation of the datasets of the positive results experiments listed above in terms of the ``standard'' SI contact interaction as follows:

\begin{itemize}
\item[$\diamond$]
The \DAMA\ modulation signal effect has been shown to be compatible with two areas in the ($m_\chi,\sigma_{\rm SI}$) parameter space, due to the different kinematics experienced by the two targets in the scintillator. In particular, a region pointing towards DM masses in the 10 GeV ballpark and towards a cross section $\sigma_{\rm SI}\simeq 2\times 10^{-40}$ cm$^2$ can be associated with a DM particle scattering off with sodium (see e.g.~the red contour in the top left-handed plot  of Fig.~\ref{fig:2} denoted by DAMA/Na).  Another region around DM masses of roughly 60 GeV is instead due to the scattering with the iodine. In this case the  cross section is enhanced by a larger  coherent factor and indeed the favored value is $\sigma_{\rm SI}\simeq  10^{-41}$ cm$^2$ (red contour denoted by DAMA/I).

\item[$\diamond$]
The modulation in \CoGeNT\ is mostly concentrated in the first two energy bins, close to the lower threshold of the detector. If this anomaly  is interpreted in terms of SI contact interaction, it will be fitted by light DM candidates (around 9 GeV) with a total SI cross section $\sigma_{\rm SI}\simeq 3\times 10^{-41}$ cm$^2$ (see e.g.~the red contour in the top right-handed plot of Fig.~\ref{fig:2}).  

\item[$\diamond$]
Like \DAMA, the \CRESST\ experiment is a multiple target detector and therefore more than one allowed region is in general expected. In particular, as one can see in Fig.~\ref{fig:2}, the region pointing towards light DM masses (around 12 GeV) and $\sigma_{\rm SI}\simeq 5\times 10^{-41}$ cm$^2$  is  due to the scattering with Oxygen and Calcium. On the other hand, the one pointing towards heavier DM mass (around 30 GeV) and cross section $\sigma_{\rm SI}\simeq 10^{-42}$ cm$^2$ can be associated with a DM particle scatters off with tungstenum. These regions are for instance denoted as gray contours in the top left-handed plot of Fig.~\ref{fig:2}.

\item[$\diamond$]
Very recently, also the 11 Silicon detectors of the \CDMS\ experiment reported an excess in their counting rate. Again, if the anomaly is interpreted in terms of SI contact interaction, it will be fitted by light DM particles (around 9 GeV) with a SI total cross section $\sigma_{\rm SI}\simeq 2\times 10^{-41}$ cm$^2 $ (blue regions in the bottom left-handed plot of Fig.~\ref{fig:2}). 
\end{itemize} 

The interesting feature is that all the experiments employing light target nuclei, seem to be compatible with a DM interpretation, which, for the ``standard'' SI contact interaction, pin-points the properties of the DM particle quite precisely; namely it leads to DM masses in the 10 GeV ballpark and total cross section in the range ($2\times 10^{-41} $ cm$^2$ $ - \, 2\times 10^{-40} $ cm$^2$). 

On the other hand, the constraints coming from null results are very stringent, and in particular the \XENONhundred\  and recently the \LUX\ results severely exclude the allowed parameter space of the positive results experiments. On a more specific level, before \LUX, the \XENONhundred\ experiment provided the most stringent constraints for a DM mass above roughly 8 GeV, with a minimum of $\sigma_{\rm SI}\simeq 1.8\times 10^{-44}$ cm$^2$ (see e.g.~the thick blue line in the top left-handed plot of Fig.~\ref{fig:2}). Nowadays this bound has been beat out by the  one of \LUX\ by a factor $\sim2.5$ (see the blue line in the bottom right-handed plot of Fig.~\ref{fig:2}). Other relevant analyses can be found in Refs. \cite{Schwetz:2011xm,Farina:2011pw,Fox:2011px,Hooper:2011hd,Gondolo:2011eq,DelNobile:2011je,Arina:2011si,Frandsen:2011ts,Feng:2011vu,Hooper:2010uy,Chang:2010yk,Fitzpatrick:2010em,Kopp:2009qt}.

\subsection{Uncertainties}
Nevertheless, since the actual range of masses and cross section  critically depends on several assumptions,  one always bears in mind all the uncertainties that enter in this field when interpreting the data in terms of a specific DM interactions.  

\begin{itemize}
\item
A first class of uncertainties comes from the poor knowledge of the properties of the DM halo. The often used Maxwell-Boltzmann halo with isotropic velocity dispersion is only a benchmark choice, not the physical description of it. Other possible choices (spherical halo with non-isotropic velocity dispersion, Axisymmetric halo, Triaxial halo and the possibility to have co-rotating halo with different number density) must be taken into account, in order to have a much better idea of how the velocity distribution affects the final results. In particular,  it has been recently demonstrated that the  uncertainties on the astrophysical part of the rate  can be removed by comparing different experimental results in the $v_{\rm min}$ space (see e.g. Refs.~\cite{Fox:2010bz,Fox:2010bu, McCabe:2011sr, Frandsen:2011gi, HerreroGarcia:2011aa, HerreroGarcia:2012fu, Gondolo:2012rs, Frandsen:2013cna, DelNobile:2013cta}). In this way one can for instance relax a bit the tension between the positive results and the constraints  showed in Fig.~\ref{fig:2} up. However,  this is not still sufficient to reconcile the complicated puzzle the experimental data have left to us (see e.g. Refs.~\cite{DelNobile:2013cta, Bozorgnia:2013hsa, DelNobile:2013cva, DelNobile:2013gba}).

\item
A second class of uncertainties comes from the experimental side.  For instance, the direct measurements of quenching factor are performed with reference detectors, and thus the systematic uncertainties, important for all the detectors in direct DM searches, could be larger than what we normally assumed. 
These must be carefully investigated, because they produce a significant  shift of the allowed regions in the Total cross section/DM mass plane. 

\item
Together with these main uncertainties, the third and perhaps the most important class of them  comes from the nature of the DM interaction, and in turn from the nuclear responses of the target nuclei. Most of the models tested in recent years are based on the assumption that the mechanism of interaction is realized through a contact interaction. Deviations from this standard phenomenological approach are interesting to study, since, from more general conditions, the DM parameter space can be analyzed. In particular, there is a concrete possibility to remove the uncertainties coming from the nature of interaction thanks to the formalism of non-relativistic operator reviewed in Sec.~\ref{NRformalism}. On a more specific level, if the experimental collaborations, in addition to present the results in terms of a specific DM-nucleus interaction, also provide all the model independent \emph{integrated form factors}, it will be  extremely useful for the community.  Indeed, in this way  one can easily compute the expected number of events for any kind of interactions (e.g.~including isospin violating interactions, momentum-dependent form factors, velocity dependent form factors) and  compare it directly with the experimental results. 

\end{itemize}

Therefore the data must be treated very prudently with a maximally conservative attitude, simply because slightly modification of the assumptions summarized above can affect the theoretical interpretation of the experimental results in a relevant way. In the next section, I will focus on the uncertainties coming from the nature of interaction and in particular I'm going to show how the allowed regions and constraints are modified, if an exchanged momentum dependent DM-nucleus interaction is taken into account.

\section{Long-Range Interaction}\label{LRInteraction}

So far, the \LHC\ has not reported any evidences of physics beyond the standard model. An optimistic point of view is that the new physics threshold is truly around the corner. A pessimistic and maybe more realistic point of view is that  the \LHC\ results are instead telling us that the TeV scale is not a fundamental energy scale for new physics and therefore all the new theories beyond the standard model are, let's say, unnatural. Within this picture there is not any particular reason to expect new particles in the TeV mass range weakly interacting with ordinary matter. Therefore it should be time to ask ourselves whether there are other indications of the relevant energy scale. DM may play a central role in this picture. Indeed, the closeness between $\Omega_\chi$ and $\Omega_{\rm b}$, usually referred as a cosmic coincidence problem, might suggest a profound similarity between the dark and the ordinary sectors. Indeed, both $\rho_\chi$ and $\rho_{\rm b}$ scale as $1/a^{3}$ with the expansion of the Universe, and their ratio is independent on time. Why  these two fractions are then comparable, if the ordinary and the dark sectors have a drastically different nature and different origin? An hidden parallel sector (mirror world\footnote{The idea of a mirror world was suggested before the advent of the Standard Model (see e.g. Refs.~\cite{LeeYang,KobzarevOkunPomeranchuk}). The idea that the mirror particles might constitute the DM of the Universe was instead discussed in Refs.~\cite{Blinnikov:1982eh,Blinnikov:1983gh}.}) may shed light on this cosmic coincident problem \cite{Berezhiani:2006ac}. Indeed, it is tantalizing to imagine that the dark world could be similarly complex (CP-violating and asymmetric), full of forces (e.g. dark electromagnetism), and matter (dominant constituent with a mass around few GeV) that are invisible to us. For a review of mirror dark world see e.g.~Refs.~\cite{Foot:2004pa,Berezhiani:2005ek}. 

\medskip
To be more concrete the phenomenology of a complex dark sector, in which the matter fields are charged under an extra $U(1)$ gauge group, is particularly interesting. In this case,  the physics of the dark sector in itself can be as complicated as the one of our sector, providing at the same time also a feeble interaction between the two worlds. Indeed,  thanks to the following renormalizable Lagrangian
\beq
\Lag=\frac{\epsilon_\phi}2 F^{\mu\nu} F_{\mu\nu}' \ ,
\eeq
 the new $U(1)$ gauge boson $\phi$ may possess a small kinetic mixing $\epsilon_\phi$ with the ordinary photon. Here $F_{\mu\nu}$ and $F_{\mu\nu}'$ are the field strength tensors of the ordinary and  ``dark'' electromagnetism respectively. One effect of this mixing is to cause DM particles to couple with ordinary particles with effective milli-charge $\epsilon_\phi e$ \cite{Holdom:1985ag, FootLewVolkas, Foot:2000vy} ,and therefore  a Rutherford like interaction gives rise if the mass of the new $U(1)$ gauge boson is smaller than the typical exchanged momentum in the interaction.

\subsection{Derivation of the main equations}
A part from the theoretical motivations which are interesting and quite strong, what describes in the non-relativistic limit the scattering between milli-charged DM particles and the protons\footnote{Notice that  the interaction here maximally violates the isospin, e.g.~the DM-neutron coupling is zero.} is the following Yukawa potential 
\beq\label{Vr}
V(r) = \frac{\epsilon_\phi}{4\pi}\frac{e\, e_\chi}r \,e^{-m_\phi r},
\eeq
where $\epsilon_\phi e$ is the effective milli-charge of the protons inside the nucleus  felt by a DM particle with ``dark'' charge $e_\chi$. Here $m_\phi$ is the mass of the dark photon that acts like an electronic cloud which screens the charges of the particles involved in the scattering.  The DM-proton matrix element can be then obtained by performing the Fourier transform of Eq.~\eqref{Vr} which writes
\beq
\Mel_p=4 \,m_p m_\chi \int_0^\infty \ud r \, 2\pi r^2 \int_{-1}^{+1} \ud \cos\theta\,e^{-i\,q\,r\,\cos\theta}\,V(r)=\mathfrak{c}_Y^p(q^2)\,\Op_1^\NR \ ,
\eeq
where $q=(2 m_\mathcal N \ER)^{1/2}$ is the exchanged momentum, $\theta$ is the scatter angle in the centre-of-mass frame and $\mathfrak{c}_Y^p(q^2)=4 \,\epsilon_\phi e\,e_\chi m_p m_\chi/(q^2+m_\phi^2)$. Since $\Op^\NR_1=\unop$, the interaction, as   already mentioned in Sec.~\ref{NRformalism}, is SI like the usual case but with a coefficient which is instead dependent on the exchanged momentum  in the scattering. Summing now over the total number of protons in the nucleus $Z_{\cal N}$ and,  if the dark sector is made of stable composite particles, over the total number of charge elementary constituents $Z_\chi$, the square of the DM-nucleus matrix element  reads
\beq
\overline{\left| \Mel_{\cal N} \right|^2} =16 \,m_\mathcal N^2 m_\chi^2 \left(\frac{\epsilon_\phi \,Z_{\cal N}e\,Z_\chi e_\chi}{q^2+m_\phi^2}\right)^2 F_{\rm Helm}^2(q^2) \ ,
\eeq
where $F_{\rm Helm}^2(q^2)$ is the usual Helm form factor. Notice that in general this equation should be also multiplied for the form factor of the composite DM particle. For our computation we assume it to be equal to one. Plugging back the DM-nucleus matrix element in Eq.~\eqref{sigmaT} the differential cross section will be
\beq\label{sigmaTlr}
\frac{\ud \sigma_{\mathcal N}}{\ud \ER}=\frac{8\pi\,m_{\mathcal N}}{v^2} \epsilon_\phi^2 \alpha^2 k_\chi^2 \left(\frac{1}{q^2+m_\phi^2}\right)^2 Z_{\mathcal N}^2 F_{\rm Helm}^2(q^2) \ ,
\eeq
where $\alpha=e^2/4\pi$ is the electromagnetic fine structure constant and $k_\chi=Z_\chi\,e_\chi/e$ is a factor that measures the strength of the DM-dark photon coupling. 

In order to have a rough comparison with the ``standard'' picture (SI contact interaction), the differential rate of nuclear recoil can be cast in the following form
\beq\label{RateSIlr}
\frac{\ud R_{\cal N}}{\ud \ER} =\frac{\xi_{\cal N}}{m_{\cal N}} \frac{\rho_\odot}{m_\chi} \left[\frac{m_{\mathcal N}}{2 \mu_N^2} \,A_{\mathcal N}^2\,\sigma_{\rm \phi\gamma}^p \,   \mathcal I (\ER)\, F^2_{\rm Helm}(\ER)\right] \mathcal{G}(\ER) \ ,
\eeq
where
\beq\label{sigmaphigamma}
\sigma_{\phi\gamma}^p=\frac{16\pi\,\alpha^2 \, \epsilon_\phi^2\,k_\chi^2}{m_N^4}\mu_N^2\simeq \left(\frac{\epsilon_\phi}{10^{-5}}\right)^2\left(\frac {k_\chi}1\right)^2\left(\frac{\mu_N}{1\mbox{ GeV}}\right)^210^{-40}\mbox{ cm}^2
\eeq
is a normalized total cross section that encapsulates all the dependences on the dark sector vertex and the kinetic mixing parameter, and the function 
\beq\label{GER}
\mathcal G(\ER)=\frac{Z_{\mathcal N}^2}{A_{\mathcal N}^2}\left[\frac{m_N^2}{q^2+m_\phi^2}\right]^2=\frac{Z_{\mathcal N}^2}{A_{\mathcal N}^2}\left[\frac{m_N^2}{2 m_{\mathcal N}\ER+m_\phi^2}\right]^2 \ 
\eeq
measures the deviation of the allowed regions and constraints compared to the ``standard'' SI contact cross section. Here the function $\mathcal G(\ER)$ is a sort of DM form factor which clearly exhibits two limits: 
\begin{itemize}
\item[$\diamond$] 
Point--like limit ($q\ll m_\phi$): in this regime $\mathcal{G}(\ER)=(Z_{\mathcal N}/A_{\mathcal N})^2   m_N^4/ m_\phi^4$ is independent on $E_R$ and therefore the interaction is of a contact type. Indeed, the rate of nuclear recoil turns out to be proportional to $\alpha^2 \epsilon_\phi^2 k_\chi^2/m_\phi^4$, which plays the same role of the Fermi's constant in weak interaction. The factor $(Z_{\mathcal N}/A_{\mathcal N})^2$ shows up the fact that milli-charged DM particles only couple with protons.   Therefore it is expected that the allowed regions and bounds in the plane ($m_{\chi},\sigma_{\phi\gamma}^p$) will be rigid shifted of a factor $\sim4$ up with respect to the ``standard'' SI scenario, in which a DM particle symmetrically couples both with neutrons and protons.

\item[$\diamond$] 
Long--range limit ($q\gg m_\phi$): in this regime $\mathcal{G}(\ER)=(Z_{\mathcal N}/A_{\mathcal N})^2   m_N^4/(4 m_{\mathcal N}^2\ER^2)$ and therefore the differential cross section acquires an explicit dependence on the nuclear recoil energy, and a Rutherford-like cross section gives rise ($\ud \sigma_{\mathcal N}/\ud\ER \propto 1/q^4$). Experiments with low energy thresholds and light target nuclei (e.g.~\DAMA\ and \CoGeNT) are therefore more sensitive than the ones with high thresholds and heavy targets (e.g.~\XENONhundred). The compatibility among the experiments could therefore be improved. 

\end{itemize}

Considering typical nuclei ($m_{\mathcal N}\sim 100$ GeV) and recoil energies (few keV) in the range of interest of the current experiments, the transition between the two limits is obtained for $m_\phi \sim \mathcal{O} (10)$ MeV. Since $q^2\propto m_{\mathcal N}$, increasing $m_{\mathcal N}$ the transition occurs at lower $m_\phi$. Notice that, once the long-range regime is reached ($m_\phi\lesssim 10$ MeV), the differential cross section is independent on the mass of the mediator. 

\begin{figure}[t]
\includegraphics[width=.325\textwidth]{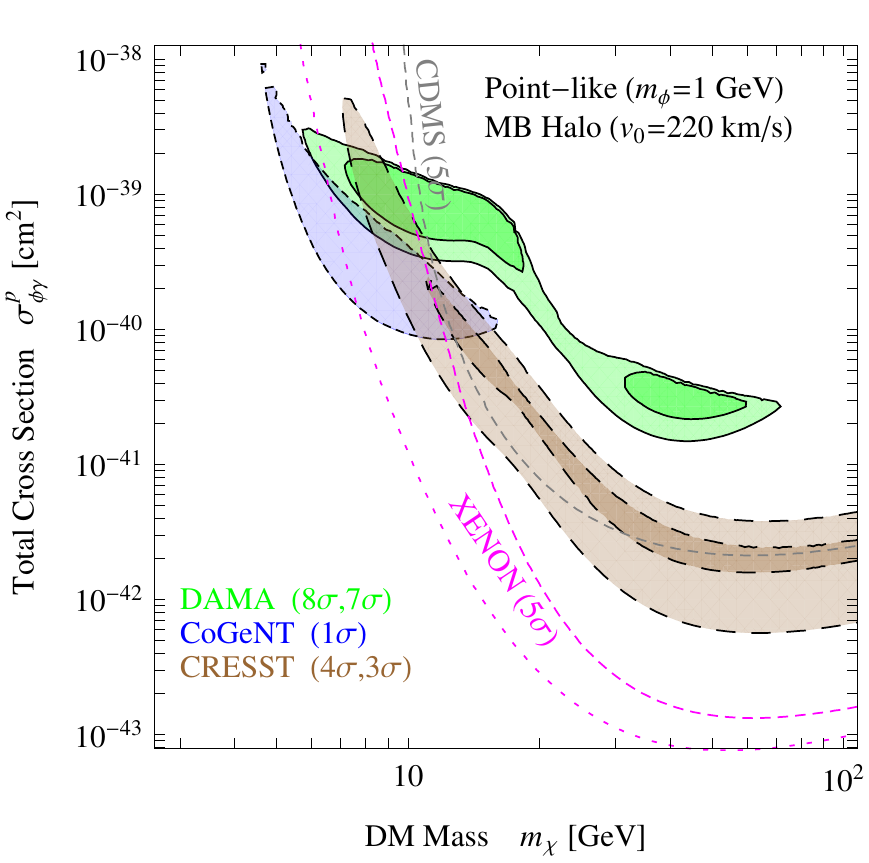}
\includegraphics[width=.325\textwidth]{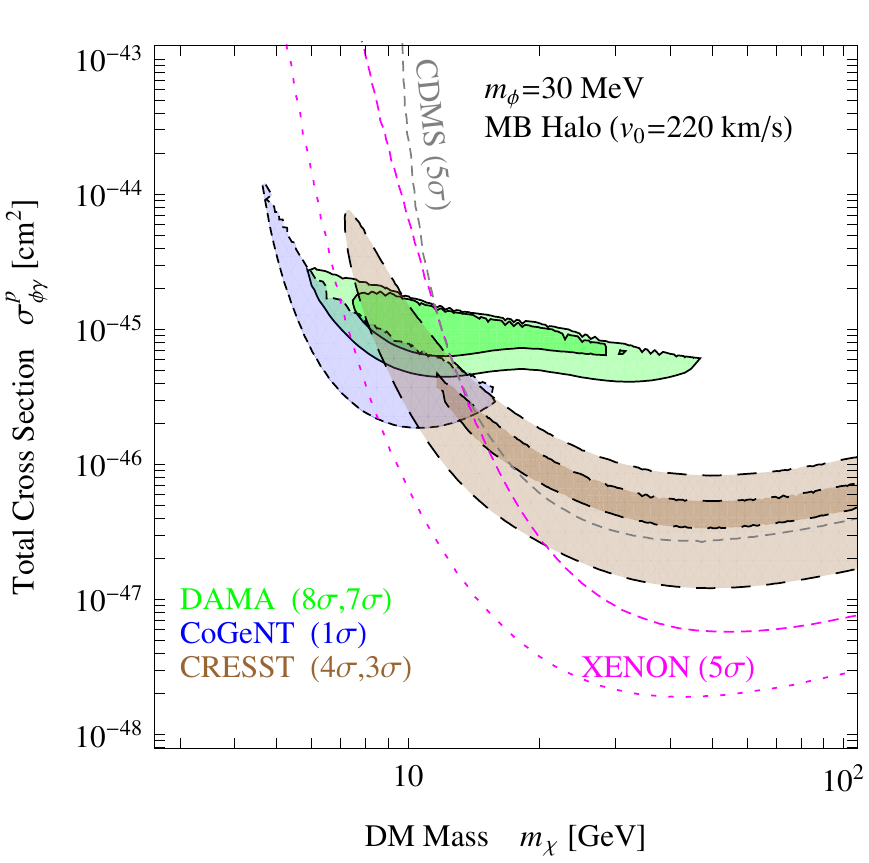}
\includegraphics[width=.325\textwidth]{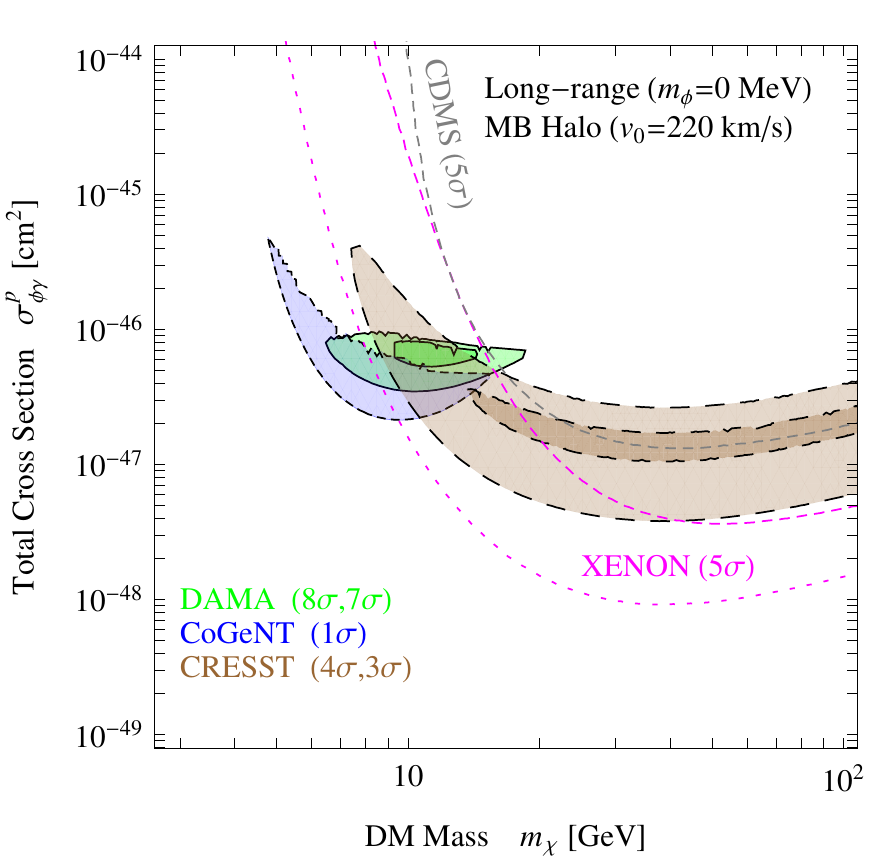}
\caption{\em \small \label{fig:3} 
{\bfseries Normalized total cross section as a function of the dark matter mass}. For a better visualization of the transition from the contact to the long-range regime, we have been fixed three values of the dark photon mass; namely $m_\phi=1$ GeV (left plot), $m_\phi=30$ MeV (central plot) and $m_\phi=0$ (right plot). In all plots we show the regions where the absence of annual modulation can be rejected at 7$\sigma$ CL (outer region) and 8$\sigma$ CL (inner region) for \DAMA\ (solid green contours), and 1$\sigma$ CL for \CoGeNT\ (solid green contours). For \CRESST\ (solid green contours) the absence of an excess is instead excluded at 3$\sigma$ CL (outer region) and 4$\sigma$ CL (inner region). Constraints, derived at 5$\sigma$ CL, are shown as dashed gray lines  for \CDMSGe\ and  magenta lines for \XENONhundred. These results have been presented in Refs.~\cite{Fornengo:2011sz,Panci:2012qf}.}
\end{figure}

\subsection{Results}
In our analysis we  consider as free parameters the normalized total cross section $\sigma_{\phi\gamma}^p$, the mass of the dark photon $m_\phi$ and of course the one of the DM particle $m_\chi$. The velocity distribution has been assumed to be Maxwell-Boltzmann-like with a velocity dispersion $v_0=220$ km/s. For the local DM density we have choosen $\rho_\odot=0.3$ GeV/cm$^3$.  For all the positive results  experiments  we use a different technique for analyzing the datasets respect to the one often found in literature.  Specifically, we adopt the same approach of \cite{Belli:2011kw}, in which the null hypothesis is tested. From this kind of statistical test we can then extract the level at which the absence of signal on top of the background in \CRESST\ and the absence of modulation in \DAMA\ and \CoGeNT\ is rejected.  

\medskip
In Fig.~\ref{fig:3} the allowed regions of the positive results experiments and constraints coming from null results in the plane ($m_{\chi},\sigma_{\phi\gamma}^p$) are shown. For a better visualization of the transition from the contact to the long-range regime, we have fixed three values of the dark photon mass; namely $m_\phi=1$ GeV (left plot corresponding to the contact limit), $m_\phi=30$ MeV (central plot) and $m_\phi=0$ (right plot corresponding to the pure long-range regime). In all plots the  solid green contours individuate the regions compatible with the \DAMA\ annual modulation  \cite{Bernabei:2008yi,Bernabei:2010mq}, without considering the channeling effect. The short-dashed blue contour refers to the region derived from the \CoGeNT\ annual modulation signal data published in \cite{Aalseth:2011wp}\footnote{In Refs.~\cite{Aalseth:2014eft,Aalseth:2014jpa}  the collaboration re-analyzed the data finding that the background of surface events was underestimated. In view of that we expect that the \CoGeNT\ allowed regions  in the ($m_\chi,\sigma_{\phi\gamma}^p$) plane will shrink around its best value, like the one reported for the ``standard'' SI contact interaction (see e.g. the red contour in the top right-handed plot of Fig.~\ref{fig:2}).}. 
The dashed brown contours individuate the allowed regions compatible with the \CRESST\ excess \cite{Angloher:2011uu}. For each experiment the contour lines cover at least one area in the   ($m_{\chi},\sigma_{\phi\gamma}^p$) plane. Specifically, they refer to regions where the absence of annual modulation can be rejected at 7$\sigma$ CL (outer region) and 8$\sigma$ CL  (inner region) for \DAMA, and 1$\sigma$ CL for \CoGeNT. For \CRESST\ the absence of an excess is instead excluded at 3$\sigma$ CL (outer region) and 4$\sigma$ CL (inner region).

Very recently, also the 11 Silicon detectors of the \CDMS\ experiment have reported an excess in their counting rate. Since the results presented in Fig.~\ref{fig:3} are based on Ref.~\cite{Fornengo:2011sz} published in 2011, we do not include a full statistical analysis of the \CDMSSi\ data-sets.  
We attempt however an analytical comparison between the results arising from the interaction studied here and the standard SI contact picture (blue regions  in the bottom left-handed plot of Fig.~\ref{fig:2}). 
More precisely, defining $\langle \mathcal G  \rangle \equiv \mathcal G(\langle \ER \rangle)$, where $\langle \ER \rangle\simeq 10$ keV is the average recoil energy of the three events observed \cite{Agnese:2013rvf}, the \CDMSSi\ allowed region in the ($m_\chi, \sigma_{\rm SI}$) plane can be projected in the parameter space of Fig.~\ref{fig:3}, through the relation $(m_{\chi},\sigma_{\phi\gamma}^p)\equiv (m_{\chi},\sigma_{\rm SI}/\langle \cal G \rangle)$. Recalling now that the best fit value of the SI cross section is $\sigma_{\rm SI}^{\rm best} \simeq 2\times 10^{-41}$ cm$^2$, we expect that the \CDMSSi\ allowed region in the ($m_{\chi},\sigma_{\phi\gamma}^p$) plane will still point towards light DM candidate, but with a normalized total cross section  $\sigma_{\phi\gamma}^p=A_{\rm Si}^2/Z_{\rm Si}^2 m_\phi^4/m_N^4 \, \sigma_{\rm SI}^{\rm best}\simeq 10^{-40}$ cm$^2$ in the point-like regime ($m_\phi=1$ GeV), and $\sigma_{\phi\gamma}^p=A_{\rm Si}^2/Z_{\rm Si}^2 4 m_{\rm Si}^2 \langle\ER \rangle^2/m_N^4 \, \sigma_{\rm SI}^{\rm best}\simeq 3\times 10^{-47}$ cm$^2$ in the long range one ($m_\phi=0$). 

%
%
%
%
%
%
%
%
%
%
%

\medskip
Constraints, derived at 5$\sigma$ CL, are shown as dashed gray lines  for \CDMSGe\ and  magenta lines for \XENONhundred.  In particular, in order to bracket as much as possible the uncertainties coming from the poor knowledge of the detection efficiency close to the lower threshold, for the \XENONhundred\ bound we adopt two approaches: $i)$ a case (dotted lines) in which the constraints are computed with the nominal value of the lower threshold of 4 PHE  and of the effective luminosity \cite{Aprile:2011hi}, which  is a sort of quenching factor in liquid xenon; $ii)$ another case (dashed lines) in which the constraints are computed by considering an higher threshold of 8 PHE: this is in order to determine a situation which is nearly independent on the poor knowledge of the effective luminosity (especially at very low nuclear recoil), and more important on the statistical distribution of the few PHE collected close to the threshold of the detector. For similar discussions see Refs.~\cite{Bernabei:2008jm,Collar:2011wq}.

In October 2013, also the \LUX\ collaboration announced their first DM search results.   With the data collected in just 85 live-days, they were able to set the most stringent bound on the SI contact cross section, with respect to the preexisting limits. We do not perform a full statistical analysis of their data, since, as already pointed out, the results in Fig.~\ref{fig:3} are taken from Ref.~\cite{Fornengo:2011sz}. Nevertheless, since both \XENON\ and \LUX\ are based on the same double-phase xenon technology, we can estimate the bound on  $\sigma_{\phi\gamma}^p$, just by rescaling the magenta lines in Fig.~\ref{fig:3} with the exposures of the two experiments.  Naively, we then expect that the \LUX\ bound will be a factor $w_{\rm \LUX}/w_{\XENONhundred}\simeq 1.5$ more stringent with respect to \XENONhundred.

\begin{figure}[t]
\centering
\includegraphics[width=.475\textwidth]{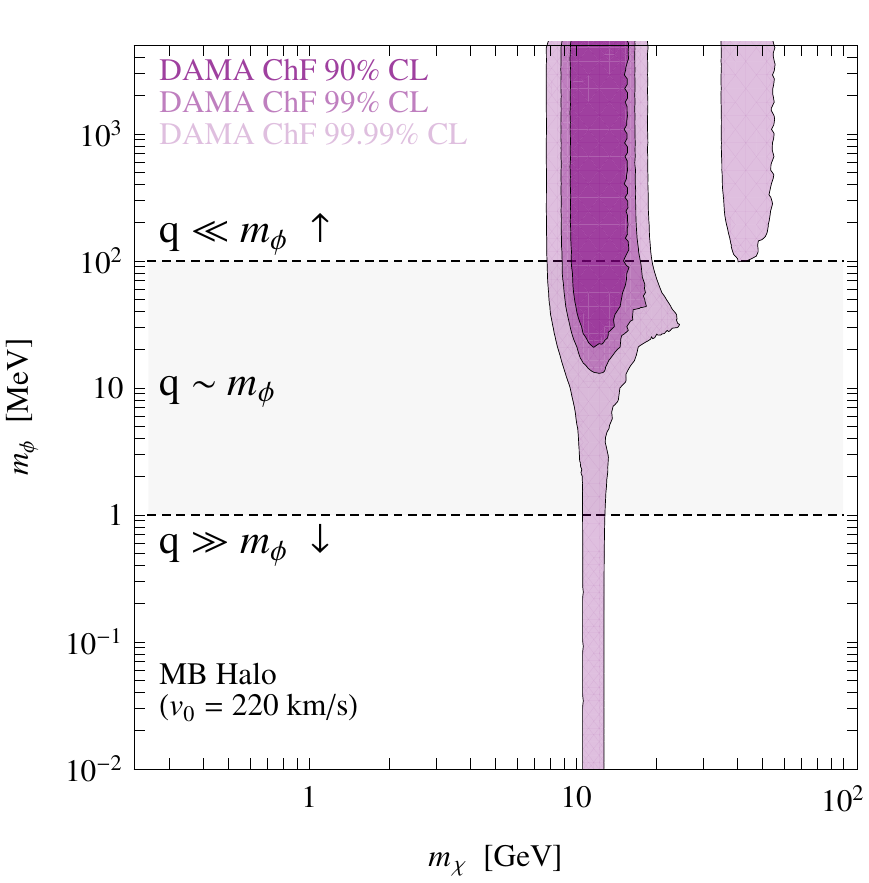}\quad
\includegraphics[width=.475\textwidth]{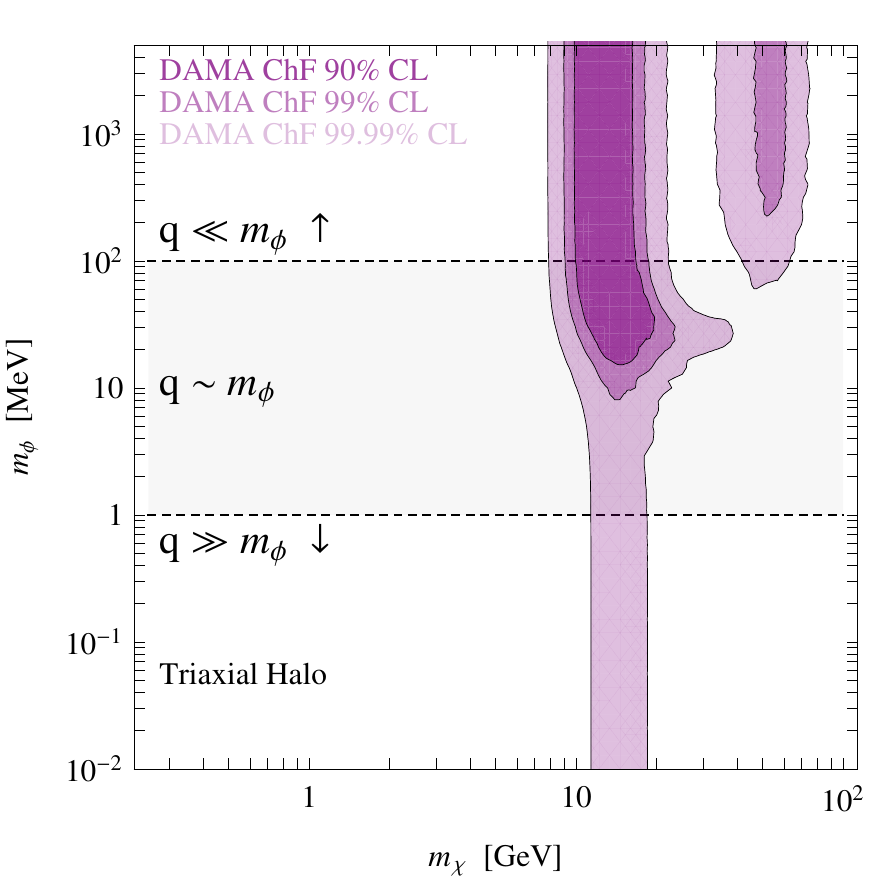}
\caption{\em \small \label{fig:4} 
The {\bfseries bounds on the mediator mass as a function of the DM one} obtained by comparing the \DAMA\  modulated signal with the total rate. The allowed regions are reported for CL's of 90\%, 99\% and 99.99\% in shaded magenta. In the left plot an isothermal sphere with velocity dispersion $v_0=220$ km/s and a local DM energy density $\rho_\odot=0.3$ GeV/cm$^3$ has been assumed. In the right plot we instead show  the allowed regions by considering a more physical triaxial halo with the same velocity dispersion in the major axis and with $\rho_\odot=0.84$ GeV/cm$^3$. These results have been presented in Refs.~\cite{Fornengo:2011sz,Panci:2012qf}.}
\end{figure}

\medskip
We can see, as expected from the discussion above, that the agreement among the positive results experiments increases moving towards long range interaction without being excluded by both the \XENONhundred\ and \LUX\ bounds, if our conservative choice for the lower threshold is taken into account.  In particular, a common region pointing towards a DM mass around  15 GeV  with a normalized total cross section $\sigma_{\phi\gamma}^p \sim  10^{-46}$ cm$^2$ has been found. Since the normalized total cross section $\sigma_{\phi\gamma}^p$ depends collectively on the kinetic mixing parameter $\epsilon_\phi$ and on the DM-dark photon coupling encoded in the parameter $k_\chi$ (see Eq.~\eqref{sigmaphigamma}), keeping fixed for instance two values of $k_\chi$, namely $k_\chi=(1,10)$, the best fit  for the kinetic mixing parameter would then be $\epsilon_\phi\sim(10^{-8},10^{-9})$.

\medskip
On the other hand, the significance of the \DAMA\ region alone gets lower, due to the fact that for pure long range interaction ($q\gg m_\phi$) the $1/q^4$ drop-off of the unmodulated signal rapidly overshoots the measured total rate (Fig.~1 of Ref.~\cite{Bernabei:2008yi}), that of course we treat as an intrinsic constraint in our analysis. This can be appreciated in more details in Fig.~\ref{fig:4}, where again the \DAMA\ allowed regions are shown, but in the ($m_\chi, m_\phi$) plane. As one can see, a 99\% confidence level lower bound on the mass of the dark photon around 10 MeV is possible to be placed. This is due to the fact that the unmodulated signal, unlike the modulated one, is particularly sensitive to the energy drop-off of the cross section, which is typical for Rutherford-like scattering. The situation does not dramatically vary  if we change the properties of the dark matter halo. Indeed, if we consider a more physical triaxial halo with the same velocity dispersion in the major axis (right plot of Fig.~\ref{fig:4}), the allowed region are only slightly enhanced. 

Nevertheless,  it is worth noticing that in a dark  world  filled by more than one species in thermal equilibrium, the DM particles responsible for the annual modulation  may possess  smaller velocity dispersion. 
For instance, in the context of mirror DM, the dominant species can be the mirror hydrogen, like our world, while the mass of the DM particles favored by direct detection experiments is around 16 GeV (mirror oxigen). If they are in thermal equilibrium, the relation $v_0=1/\sqrt2\,(A_{\rm H}/A_{\rm O})^{1/2}\cdot 220$ km/s $\lesssim 100$ km/s then will hold \cite{Foot:2008nw}. In this case the constraints in Fig.~\ref{fig:4} are less important, because the raise of the total rate for low nuclear recoil energy is under threshold (see e.g.~Figs.~5(a-b) in Ref.~\cite{Foot:2008nw}). 

\medskip
Since, as already stated, the normalized total cross section, and in turn the rate, depend collectively on the kinetic mixing parameter $\epsilon_\phi$ and on the DM-dark photon coupling $k_\chi$, in the next section, a complementary class of constraints which are relevant for DM models featured by a long range interaction will be presented. 

\subsection{Complementary Constraints}

\begin{figure}[t]
\centering
\includegraphics[width=.475\textwidth]{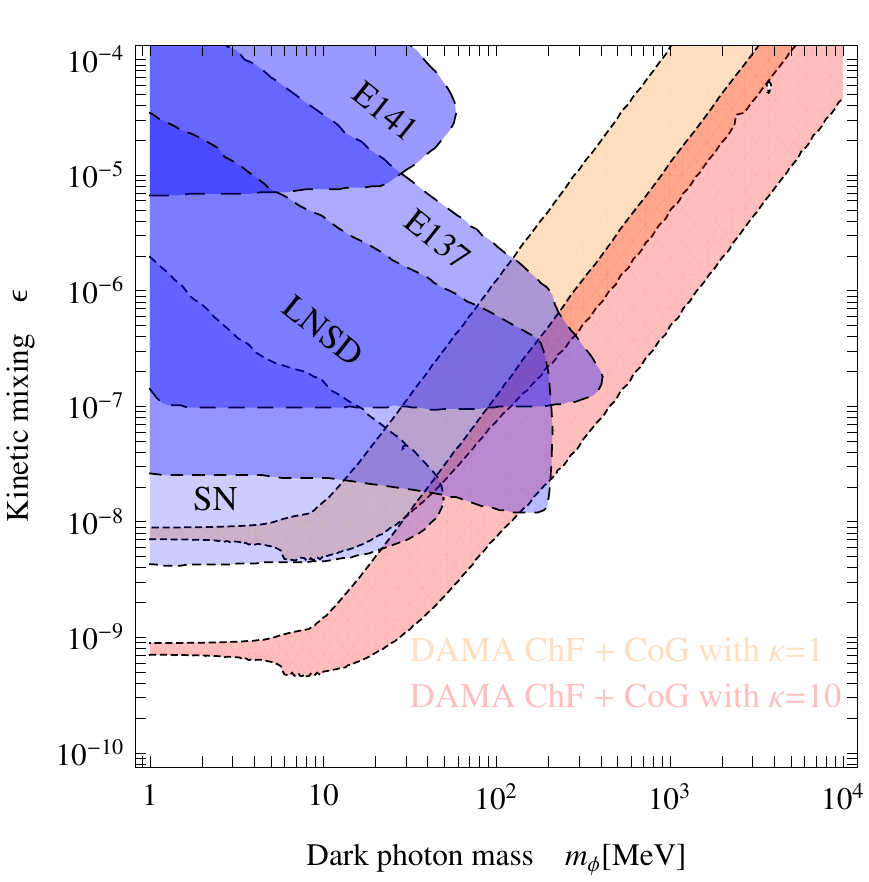}\quad
\includegraphics[width=.475\textwidth]{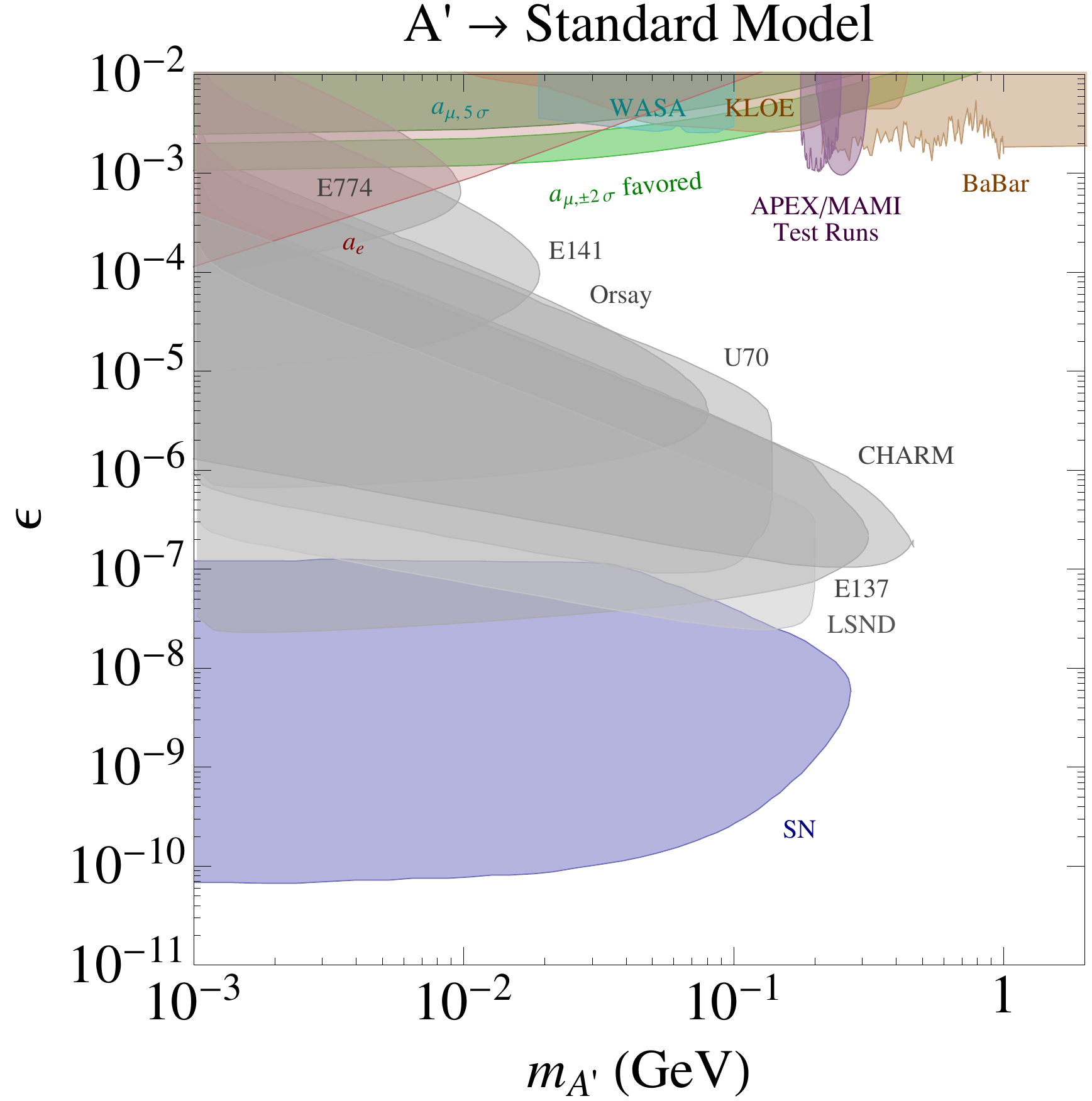}\quad
\caption{\em \small \label{fig:5} 
{\bfseries Allowed regions and a compilation of constraints} in the $(m_\phi,\epsilon_\phi)$ plane. Left plot: The upper (orange) and the lower (red) allowed regions are derived by combining the \DAMA\ and \CoGeNT\ annual modulation dataset. Specifically, the absence of modulation  is here rejected at 8$\sigma$ CL. The two regions differ in the choice of the parameter $k_\chi$, namely $k_\chi=1$ (upper) and $k_\chi=10$ (lower). In blue are instead represented the constraints coming from beam dumped neutrino experiments and supernov\ae\ observations taken from Ref.~\cite{Bjorken:2009mm,Essig:2010gu}. Right plot: A compilation of the latest constraints in the  $(m_\phi,\epsilon_\phi)$ plane. This plot has been taken from the following review \cite{Essig:2013lka}. It is worth pointing out that for $m_\phi\lesssim 10^{-6}$ eV the parameter space of the positive results experiments in direct DM searches is basically unbounded (see e.g. Fig.~7 in Ref.~\cite{Essig:2013lka})}
\end{figure}

A first class of complementary constraints that solely depends on the properties of the dark photon (kinetic mixing parameter $\epsilon_\phi$ and its mass $m_\phi$), is the one coming from beam dump neutrino experiments and supernov\ae\ observations. In Fig.~\ref{fig:5} a compilation of the bounds in the 
$(m_\phi,\epsilon_\phi)$ plane is shown. Specifically, on the left panel the shaded blue regions are taken from Ref.~\cite{Bjorken:2009mm,Essig:2010gu}, while those in the right one represent the latest constraints on the dark photon properties presented in Ref.~\cite{Essig:2013lka}\footnote{Notice that the labels $\epsilon$ and $m_{A'}$ in the plots of Fig.~\ref{fig:5} stand for the kinetic mixing parameter $\epsilon_\phi$ and the dark photon mass $m_\phi$ respectively.}. As is apparent, the bounds coming from supernov\ae\ observations (namely the energy loss observed from SN1987a) are  the most stringent. They exclude small kinetic mixing parameter and light mass for the dark photon. Indeed, in this case, if dark photon were produced in relevant amount in the centre of a supernova via the mixing with the ordinary photon, the subsequent emission of them would shorten so much the predicted neutrino boost that it would become inconsistent with the measurements made  by kamiokande  (see e.g.~Refs.~\cite{Bjorken:2009mm, Dent:2012mx, Dreiner:2013mua}). 

The allowed regions of the \DAMA\ and \CoGeNT\ experiments projected in the $(m_\phi, \epsilon_\phi)$ plane are instead shown  in the left plot of Fig.~\ref{fig:5}. Since, as already stated, the normalization of the rate depends not only on the properties of the dark photon but also on  its coupling with the DM particles, we present the results for two benchmark values of $k_\chi$, e.g.~$k_\chi=1$ (upper orange region) and $k_\chi=10$ (lower orange region).  We can see that in the ``symmetric'' case $k_\chi=1$, only the dark photons with $m_\phi\gtrsim 100$ MeV can simultaneously satisfy the constraints and give a reasonable fit of the positive results experiments. On the other hand, for $k_\chi\gtrsim 50$, the whole range is basically unbounded. It is worth stressing  that such large values of $k_\chi$ can be easily obtained either in models of composite DM particles with large $Z_\chi$ (e.g. mirror models) or in those with a strongly coupled dark sector. Furthermore, it is also relevant to point out that for pure long-range interaction ($m_\phi=0$), the majority of the bounds, and in particular those coming from supernov\ae\ observations, do not apply, being the direct production of dark photon forbidden by kinematical reason (see e.g. Fig.~7 in Ref.~\cite{Essig:2013lka}). 

\medskip
A second class of constraints which instead solely depends  on the properties on the dark sector itself is the one coming from the self-interactions. Indeed, since for this class of models the DM-DM scattering, given by 
\beq\label{Vself}
V(r)=\frac{\alpha \, k_\chi}{r}\, e^{-m_\phi \, r}, 
\eeq
is not suppressed by $\epsilon_\phi$, the self-interactions can easily reach high values, making the dynamics of virialized astrophysical objets affected. This is particularly relevant in the limit $m_\phi\rightarrow 0$; indeed, in this case the self-interaction~\eqref{Vself} could be of the same order of the electromagnetic scattering. 

A first example of such bounds comes from the observations of colliding galaxy clusters, like the bullet cluster \cite{Clowe:2006eq}, which points towards collision-less DM. In particular a quite robust bound $\langle \sigma_T \rangle/m_\chi \lesssim 1.25$ cm$^2$/g on the size of the self-interaction has been placed by \cite{Randall:2007ph}. In order to translate such bound in a constraint on the dark photon mass, one has to compute the weighted energy transfer cross section, which measures the rate at which energy is transferred in the system. It reads
\beq
\langle \sigma_T \rangle = \iint \ud^3 v_1 \ud^3 v_2 \, f(v_1)f(v_2) \, \sigma_T(v_{\rm rel}) \ ,
\eeq
where $ \sigma_T(v_{\rm rel})=\int \ud\Omega \, \ud  \sigma / \ud \Omega\,(1-\cos\theta)$ is the two-bodies energy transfer cross section. Here $f(v)$ is the DM velocity distribution, assumed to be Maxwellian, and $v_{\rm rel}=|\vec v_1-\vec v_2|$ is the relative velocity of the DM particles involved in the scattering. Considering now the typical velocity of collision in the bullet cluster of 4700 km/s, a DM mass of 10 GeV and two values of $k_\chi=(1,10)$, the bound of 1.25 cm$^2$/g is exceeded if the mass of the dark photon is smaller than (1,20) MeV. 

A second example of bounds, which is in principle relevant for this kind of models, comes from the fact that quite large self-interaction rapidly drive the DM halo into a spherical configuration, due to the prompt equipartition of the energy in the system. In particular, the efficiency of such process  can be roughly estimated by 
\beq\label{taurel}
\tau_{\rm rel}=1/\langle \Gamma_T \rangle \ , \qquad \mbox{where } \langle \Gamma_T \rangle=\iint \ud^3 v_1 \ud^3 v_2 \, f(v_1)f(v_2) \, n_\chi v_{\rm rel} \sigma_T(v_{\rm rel}) \frac{v_{\rm rel}^2}{v_0^2}\ ,
\eeq
which is telling us the typical time at which the self-interactions affect the dynamics of a virialized astrophysical objet with  number density $n_\chi$ and dispersion velocity $v_0$. Indeed, if it is smaller than the age of the object, a spherical configuration will tend to form, making such scenario excluded by few  elliptical DM halos observations \cite{Feng:2009mn, Feng:2009hw,Buckley:2009in,Ibe:2009mk,Loeb:2010gj}. In particular, in the case of galaxy clusters we get that the relaxation time is always bigger than the age of the object ($\tau_{\rm ob}\sim 10^{10}$ years), making therefore the mass of the dark photon basically unbounded. On the other hand, if we consider a smaller halo, like the one of dwarf galaxies, which are characterized by larger number density and smaller velocity dispersion, one can in principle put a very stringent lower bound on the mass of the dark photon of the oder of 100 MeV (see also Refs.~\cite{Kaplinghat:2013yxa,Tulin:2013teo}). 
  
\medskip
Nevertheless, it is worth stressing that the derivations of the constraints coming from self-interaction are affected by several uncertainties both from the theoretical and experimental side. Indeed, since the phenomenology of this class of models is completely different compare to the standard one, an N-body/hydrodynamical simulation is needed. This is  especially true in the case of multi-component dark sector, in which the different equipartition of the energy among the DM particles can generate a sort of dark electric and magnetic fields in the long range regime.  Furthermore, since the self-interaction needed to change their dynamics is in general of the order of the Thomson scattering ($\sigma_{\rm em}\sim 10^{-24}$ cm$^2$), from trivial analysis dimension of such large cross section, the following rough estimation yields:  the self-interaction is of a long range type in most of the virialized astrophysical objects under the assumption that the DM-dark photon coupling is of the order of $\alpha$. In this case,  the probability to radiate a dark photon in the scattering process is then different from zero and therefore it might well be possible that the DM sector is dissipative like ours. The time at which energy is transferred in the system ($\tau_{\rm rel}$) is no longer a good indicator, since the relevant quantity that describes the dynamical evolution of the system is now the cooling time: in particular, for a DM sector composed by heavy and light species, the dissipation time due to the soft emission of the dark photon (dark bremsstrahlung) can be smaller than the age of the virialized astro-physical objects (see e.g.~Refs~\cite{Fan:2013yva,Fan:2013tia,McCullough:2013jma}). In this scenario the system is no longer stable, and in general it starts to collapse. By virtue of this fact, we do not consider this last class of constraints, since dedicated and careful analysis involving also numerical simulations are needed.

\section{Conclusion}
Direct DM searches is now characterized by tantalizing results and hints   that make this field very active both from the theoretical and experimental side. 
In particular, in addition to the long standing \DAMA\ results, nowadays there are other experiments, like \CoGeNT, \CRESST\ and \CDMSSi\ that are starting to observe anomalies in their counting rates. On the other hand, the increasingly stringent constraints coming from null result experiments put in serious trouble the theoretical interpretation of the data, at least in terms of the simple-minded SI contact interaction. In this work I  discussed the {\em status} of direct DM detection with specific attention to the experimental results and their phenomenological interpretation in terms of DM interaction. In particular, in the first part I presented a new and more general approach to study signals in this field based on non-relativistic operators. Then I reviewed  the experimental results and their interpretation in terms of the ``standard'' SI interaction pointing out all the uncertainties which enter in this field. In the last part of this work, I investigated a fermionic Dark Matter particle carrying a small milli-charged and analyzed its impact on direct detection experiments. I showed that this kind of long range interaction can accommodate the positive experimental results. By assuming a conservative choice for the lower threshold of the \XENONhundred\ and \LUX\ experiments I have demonstrated that this candidate is not ruled out. I also determined the complementary class of constraints which are relevant for milli-charged DM particles with long range forces. 

\medskip
Finally,  I would like to propose a possible direction to pursue  in order to make sense of the current exciting experimental panorama based on the formalism of non-relativistic operators. Indeed, as we have seen in the first part of this work, it allows us to describe the DM-nucleus interactions in terms of a very limited number of relevant degrees of freedom. In this way, it is possible to parametrize the model-dependent part of the rate from the model-independent one encapsulated in a sort of \emph{integrated form factors} that encode all the dependences on the astrophysics, nuclear physics and experimental details. Therefore, since one is ignorant or agnostic about the underlying theory, I would like to encourage a synergy between nuclear physicists and experimentalists in order to provide a complete set of \emph{integrated form factors} defined in Eq.~\eqref{integratedFF}. It would be extremely useful for the community, because in this way, one can compute the expected number of events for any kind of interaction (e.g.~including isospin-violating interactions, momentum-dependent form factors, velocity-dependent form factors) and compare  directly with the experimental results. Providing the \emph{integrated form factors} will thus be  the first step towards a model-independent analysis in direct DM searches.

\small
\paragraph{Acknowledgements}
We thank Marco Cirelli, Eugenio Del Nobile and Joe Silk for useful discussions.  

\bigskip
\noindent The author declare that there is no conflict of interests regarding the publication of this article.

\bigskip
\appendix

\footnotesize
\begin{multicols}{2}
  
\end{multicols}

\end{document}